# A Far-Infrared Spectral Sequence of Galaxies: Trends and Models

Jacqueline Fischer[1], N. P. Abel[2], E. González-Alfonso[3], C. C. Dudley[4], S. Satyapal[5], and P. A. M. van Hoof[6]

## ABSTRACT

We present a framework for the interpretation of the far-infrared spectra of galaxies in which we have expanded the model parameters compared with previous work by varying the ionization parameter $U$, column density $N(H)$, and gas density at the cloud face $n(H^+)$ for a central starburst or AGN. We compare these models carried out with the *Cloudy* spectral synthesis code to trends in line-to-total far-infrared luminosity ratios, far-infrared fine-structure line ratios, IRAS colors, and OH and $H_2O$ column densities with trends found in the well-studied sample of ten nearby galaxies from the IRAS Bright Galaxy Sample with infrared luminosities greater than $10^{10}$ $L_\odot$ and IRAS 60 micron fluxes equal to or greater than that of the nearby ULIRG Arp 220. We find that the spectral sequence extending from normal starburst-type emission line spectra to ULIRG-type absorption-dominated spectra with significant absorption from excited levels, can be best explained by simultaneously increasing the hydrogen column density, from as low as $10^{21}$ cm$^{-2}$ to as high as $10^{24.8}$ cm$^{-2}$ or greater, and the ionization parameter, from as low as $10^{-4}$ to as high as 1. The starburst models best reproduce most of the sequence, while AGN models are somewhat better able to produce the high OH and $H_2O$ column densities in Arp 220. Our results suggest that the molecular ISM in ULIRG-like, absorption-dominated systems is located close to and at least partially obscures the source of power throughout much of the far-infrared, which must be taken into account in order to properly interpret diagnostics of both their sources of power and of feedback.

[1] Naval Research Laboratory, Remote Sensing Division, 4555 Overlook Ave SW, Washington D.C. 20375, USA; jackie.fischer@nrl.navy.mil
[2] MCGP Department, University of Cincinnati, Clermont College, Batavia, OH, 45103, USA
[3] Universidad de Alcalá, Departamento de Física y Matemáticas, Campus Universitario, 28871 Alcalá de Henares, Madrid, Spain
[4] Formerly at Naval Research Laboratory, Remote Sensing Division, 4555 Overlook Ave SW, Washington D.C. 20375, USA
[5] School of Physics, Astronomy, and Computational Sciences, George Mason University, MS 3F3, 4400 University Drive, Fairfax, VA 22030, USA
[6] Royal Observatory of Belgium, Ringlaan 3, 1180 Brussels, Belgium

## 1. INTRODUCTION

The development of far-infrared diagnostics to derive the conditions in local infrared-bright galaxies, which are often interacting or merging galaxies, is of great importance for the study of galaxy evolution. Via dissipative collapse, gas-rich galaxy interactions and mergers generate centrally concentrated, infrared-bright star formation as well as obscured or partially obscured nuclear accretion activity and play a critical role in galaxy evolution, being responsible for the transformation from early to late galaxy types and for feeding the accretion activity in galaxy centers (e.g. Sanders et al. 1988; Kormendy & Sanders 1992; Rothberg et al. 2013 and references therein). While spectroscopic observations of these galaxies in transition are critical for our understanding of the details of this morphological transformation, they are difficult to carry out and interpret, in large part because by their nature, infrared-bright galaxies are dusty and their power sources and environments are often obscured in the ultraviolet and sometimes throughout the infrared range. Because the effects of extinction on spectroscopic diagnostics can be varied depending on the distribution and types of dust along the lines of sight to the regions of interest, and because far-infrared spectroscopic diagnostics are less affected by extinction than at shorter wavelengths, far-infrared spectroscopic surveys of local IR-bright galaxies and careful comparison with models can provide important constraints on the power sources, chemistry, kinematics and geometry present during these evolutionary stages.

Numerous far-infrared spectroscopic studies of local galaxies based on observations obtained with the Long Wavelength Spectrometer (LWS) of the Infrared Space Observatory (ISO) and more recently with the Photoconductor Array Camera & Spectrometer (PACS) of the Herschel Space Observatory have been devoted to the application of photoionization, photo-dissociation region (PDR), and radiative transfer models to the fine-structure line and molecular absorption and emission line spectra of individual galaxies (eg. Fischer et al. 1996, Colbert et al. 1999, Bradford et al. 1999, Unger et al. 2000, González-Alfonso et al 2004, Spinoglio et al. 2005, 2012, González-Alfonso et al 2008, Fischer et al. 2010, González-Alfonso et al. 2010, 2012, 2013, 2014) and to the use of photoionization and PDR models to explain trends seen in fine-structure line emission ratios in surveys of local galaxies (eg. Malhotra et al. 2001, Negishi et al. 2001, Luhman et al 2003, Abel et al. 2009, hereafter referred to as Paper I, Graciá-Carpio et al. 2011, and Farrah et al. 2013). Luhman et al. (2003) suggested that the [CII] and other IR line deficits in ULIRGs are due to preferential absorption of ionizing photons by dust in dust-bounded ionized regions and the resulting boosted far-infrared emission from warm dust due to high ratios of ionizing photon density to hydrogen density, i.e. due to high ionization parameters. When this is the case, an effect akin to saturation occurs, as photons are not efficiently absorbed by gas at the illuminated, ionized cloud face, resulting in preferential absorption of ionizing photons by dust (e.g. Voit 1992). In Paper I, we used the *Cloudy* spectral synthesis code to model the effects of high ionization parameters in ULIRGs with visual extinction $A_V$=100 and a total hydrogen density at the illuminated face of $n(H^+)=10^3$ cm$^{-3}$ and compared our models with ISO and Spitzer Space Telescope spectroscopic observations of galaxies. We showed that the ~ 1 dex drop in the [CII]/FIR ratios, the tendency toward warmer IRAS F60/F100 ratios, and other observed properties in ULIRGs compared with lower luminosity galaxies can be qualitatively replicated by increasing the ionization parameter via an increase in the ionizing photon flux. Based on sensitive Herschel/PACS far-infrared fine-structure line measurements of ULIRGs and other galaxies, Graciá-Carpio et al. (2011) were able to extend these results, reporting on the order of magnitude deficits found for galaxies with high far-infrared luminosity-to-molecular gas mass ratio. They found that while the far-infrared line deficits observed in [CII]/FIR, [OI]/FIR, and [NII]/FIR can be explained by higher ionization parameters, the deficits observed in [OIII]/FIR and [NIII]/FIR cannot be explained by the same model parameters. Furthermore, Zhao et al. (2013) found that the submillimeter [NII] line at 205 μm is also in deficit in ULIRGs and other warm galaxies, Díaz-Santos et al. (2013) found a correlation between the [CII] deficit and the 9.7 μm silicate feature absorption strength in LIRGs, and Farrah et al. (2013) found a correlation between the silicate strength and the severity of the [CII] and [NII] far-infrared line deficits in ULIRGs and that these deficits are more severe than the deficits found for the [O I]63,145 lines. These results are consistent with the finding in the case of Mrk 231 by Fischer et al. (2010), that all IR fine-structure emission lines are in deficit compared to normal starburst and AGN galaxies with little dependence on wavelength, and that stronger deficits occur in lines with higher ionization potentials. Fischer et al. suggested that the finding in Mrk 231 could be due to partial covering of the line emitting regions, higher for high ionization lines, due for example, to a molecular torus with FIR-thick clumps.

In a series of papers modeling the ISO LWS and Herschel PACS far-infrared absorption dominated spectra of Arp 220 and Mrk 231, González-Alfonso et al. (2004, 2008) and González-Alfonso et al. (2012, 2013, 2014) showed that the observed rotational levels of OH, $H_2O$, and other molecular species are radiatively pumped by the high far-infrared radiation density present in nuclear regions of these nearby ULIRGs (see also González-Alfonso et al. 2010 for the effects on the submillimeter lines of $H_2O$). In this work, in order to investigate trends leading up to the absorption-dominated character of most local ULIRGs, we follow up the initial far-infrared spectral sequence of nearby galaxies presented by Fischer et al. (1999) by enlarging the sample from six to ten galaxies with full LWS spectra and by using the most recently available capabilities of the *Cloudy* spectral synthesis code which now includes detailed modeling of molecules in PDRs and X-ray dissociation regions (XDRs), in addition to ionized and atomic regions around starburst and AGN. We look at the differences in character found in the full far-infrared spectra of a small sample of galaxies and explore to what extent the presence of high ionization parameters, density, and/or far-infrared extinction in ULIRGs are responsible for the warm ULIRG colors, the high abundances of water and OH, and the mid- to far-infrared fine-structure line deficits.

With the recent completion of the Herschel Space Observatory mission, and in particular with the important discovery of massive, high velocity molecular outflows in ULIRGs with mid-IR evidence of luminous AGN (Fischer et al. 2010, Sturm et al. 2011, Spoon et al. 2013, Veilleux et al. 2013, González-Alfonso et al. 2014), this study of the lines throughout the full far-infrared spectroscopic range over the large ISO LWS apertures can form the basis for further detailed studies of large numbers of galaxies as the far-infrared extragalactic spectroscopic dataset expands to a much more detailed and rich characterization. Moreover, such understanding is necessary for the interpretation of present and future far-infrared and submillimeter spectroscopic observations with the Atacama Large Millimeter/submillimeter Array (ALMA), Space Infra-Red Telescope for Cosmology and Astrophysics telescope (SPICA), Cerro Chajnantor Atacama Telescope (CCAT) and beyond.

## 2. THE SAMPLE

One of the attributes of the grating mode of the ISO LWS was that due to its low spectral resolution, $R = \lambda / \Delta\lambda \approx 150 - 300$, ten simultaneously scanned detectors, and ~75" aperture (Clegg et al. 1996; Swinyard et al. 1996; Gry et al. 2003) it was capable of measuring the full 43 – 197 μm far-infrared spectra of the central regions of nearby, bright galaxies with only moderate investments of observing time, thus enabling an unbiased census of the spectral characteristics of nearby, infrared-bright galaxies. For this purpose, we have reduced and analyzed the full spectra of galaxies from the IRAS Bright Galaxy Sample (RBGS, Sanders et al. 2003) whose infrared luminosities are greater than $10^{10}$ $L_\odot$ and for which F60 ≥ F60-Arp 220, where F60 denotes the IRAS flux density, in Jansky, at 60 μm, an abbreviation used hereafter for the four IRAS bands. So defined, this sample consists of 10 nearby galaxies of a variety of morphological and nuclear types that span the range of far-infrared colors and luminosities of galaxies leading up to those of ULIRGs. The sample galaxy properties are listed in Table 1 with observed coordinates, redshifts, distances, spatial extents corresponding to the LWS aperture at the RBGS listed distance, far-infrared colors F60/F100, far-infrared and total infrared luminosities $L_{FIR}$(40–400 μm) and $L_{IR}$(8–1000 μm) from the RBGS, the galaxy type, and notes on orientation. Although variation of spatial extent due to the range in distances may be responsible for some of the differences observed, nevertheless, the spectra represent the variety of average properties sampled by the beam.

For this first comparison with our expanded model set aimed at relating the tracers of the ionized, atomic, and molecular gas in lower luminosity galaxies to those of ULIRGs, there are several reasons for adopting this sample and the associated full ISO/LWS data. For these bright galaxies, the available LWS data cover the full ~ 50 to 200 micron far-infrared spectral range, providing characterization of the strongest far-infrared molecular absorption features to probe the effects of varying ionization parameter, as well as a range of ISM column densities, while full Herschel spectra for all of this sample and full spectral range Herschel/PACS observations of a sample of more distant LIRGs do not exist. Most importantly, for these nearby galaxies, the large LWS beam (~75") provides good spatial coverage of the central regions of the nearest and thus the most extended galaxies in the sample, encompassing at least the central kiloparsec in all of the sample galaxies (Table 1, column 6) which is critical for comparison to far-infrared observations of ULIRGs, which encompass the entire bright central regions, owing to the compact nature and larger distances of these merging systems. For this reason we supplemented the LWS measurements of the sample with matching aperture ISOCAM circular variable filter (CVF) images of polycyclic aromatic hydrocarbon (PAH).

## 3. OBSERVATIONS AND DATA REDUCTION

The full LWS 43–197 μm spectra of the 10 sample galaxies were obtained with the LWS spectrometer (Clegg et al. 1996) on board ISO (Kessler et al. 1996). The grating spectral resolution was ~ 0.3 μm and 0.6 μm in the 43–93 μm and 80–197 μm ranges respectively, corresponding to velocity resolution ~ 1-2 x $10^3$ km $s^{-1}$, so the galaxy kinematic structure is unresolved. The beam full width half maximum size ranged from 65"– 80" for the 10 LWS detectors (see Gry et al. 2003 for further details on the beam profiles). ISO Target Dedicated Time (TDT) observation numbers for each galaxy in the sample are given in the notes to Table 2. The LWS grating data used for this analysis were the uniformly processed data product (HPDP) obtained from ISO Data Archive which were produced using version 10.1 of the Off Line Processing (OLP) Pipeline system (Gry et al. 2003). We performed subsequent data processing, including co-addition, scaling to produce continuous spectra, line fitting, and integration over the full LWS continuum using the ISO Spectral Analysis Package (ISAP; Sturm et al. 1998). For each sample galaxy, Table 2 lists the IRAS far-infrared fluxes calculated according to Sanders & Mirabel (1996), their Table 1, and that emitted within the LWS beam obtained by integration over the LWS full spectrum, in columns 2 and 3, as well as the line fluxes for the seven far-infrared fine-structure lines, designated by ionization state and wavelength in microns, and the equivalent widths and statistical uncertainties for the OH 119 μm doublet (unresolved with LWS, hereafter referred to as OH119).

We supplement the LWS data with PAH 6.2 μm emission feature fluxes based on ISOPHOT-S (for Arp 220) and ISO CAM CVF spectra integrated over the LWS profile-weighted beam (first presented in Luhman et al. 2003). For four of the extended galaxies (Arp 299, NGC 253, and Cen A), only ISO CAM data over a 48" x 48" image are available (smaller than the LWS aperture), but we infer from larger broadband ISO CAM images that these images include most of the flux.

In Table 3, for the sequence galaxies in which molecular lines are detected, we tabulate the molecule and transition (column 1), the rest wavelength in microns (column 2), the lower level energy, $E_{lower}$(K) (column 3), and in columns (4 – 9) split columns list continua in $10^{-15}$ W m$^{-2}$μm$^{-1}$, equivalent widths in $10^{-2}$ microns, and the equivalent width statistical uncertainties in $10^{-2}$ microns, in parentheses, for a Gaussian profile fit. A negative equivalent width indicates the line is in emission. Comparison between the measured equivalent widths and line strengths for 16 isolated OH, H$_2$O and [OI]63 doublets/lines measured with ISO/LWS (Tables 2, 3) and Herschel/PACS (González-Alfonso et al. 2012) for Arp 220 yields good agreement to within the statistical errors and expected calibration accuracies for the two instruments (10-20% for LWS and 5-10% for PACS), for all but two H$_2$O lines at 57.6 and 66.4 μm.

## 4. RESULTS AND TRENDS

Figure 1 displays the full LWS spectra of the 10 sample galaxies, plotted in decreasing order of the ([OIII]88,52 + [NIII]57) / FIR ratio as tabulated in column 13 of Table 2, hereafter referred to as ([OIII]+[NIII])/FIR and, from the bottom to the top of the figure, where FIR denotes the integrated continuum flux in the LWS aperture as listed in column 3 of Table 2. The most obvious spectroscopic trends with decreasing ([OIII]+[NIII])/FIR are that 1) the [OI]63,145, [CII]158, and [NII]122 line-to-FIR ratios remain approximately constant and 2) the molecular absorption line depths are minimal or undetected, until ([OIII]+[NIII])/FIR drops from its highest value of ~ 0.35% by nearly an order of magnitude. Hereafter we refer to the two extremes as the emission-line dominated and molecular-absorption dominated sides of the sequence. In Figure 2, we quantify these trends and others versus ([OIII]+[NIII])/FIR, showing the equivalent width of the strongest OH transition, IRAS band flux ratios, and line/FIR ratios. The ([OIII]+[NIII])/FIR error bars are based on the statistical 1-σ measurement uncertainties. In this figure the dotted lines are least-squares fits to a variety of functions – linear, polynomial, and exponential – used here solely as a visual aid. While the ratio trends show significant scatter, likely due to the differences in the galaxy types, evolutionary stages and environments of the sample galaxies, the plotted ratios show evidence for trends beyond the scatter, some monotonic and some not, which we briefly describe in this section before proceeding with the model description followed by a detailed comparison to these trends.

In Figure 2, panel a, the equivalent width of the OH119 doublet begins at low levels on the emission line side of the sequence, decreases toward the middle of the sequence and then rises again at the lowest values of ([OIII]+[NIII])/FIR. Panel b shows a similar trend for the F60/F100 ratio, which drops mid-sequence and then rises again for Arp 220, while the F25/F60 ratio shown in panel c is flat along the emission-line dominated side (with the notable outlier, NGC 1068, host to a Seyfert 2 nucleus) until it decreases at the molecular-absorption dominated end. Both the ([CII]158 + [OI]63,145) / FIR ratio, denoted hereafter as ([CII]+[OI])/FIR, and the [OI]145/FIR ratio, shown in panels d and e, remain approximately constant over most of the sequence, dropping at the molecular absorption/line deficit side. In addition to trends in line-to-FIR flux ratios, we also see trends in some line flux ratios. The [OI]63/[OI]145 ratio is approximately constant over most of the sequence though somewhat high mid-sequence, but drops for the deficit galaxies, in fact appearing in absorption in Arp 220, whose ratio is not included in panel f. The [OI]145/[CII]158 ratio is also fairly constant, with some scatter, over most of the sequence. The LWS upper limit on the OI145 line is consistent with a rise at the deficit end of the sequence, while the ratio of 0.084±0.011 measured with Herschel/PACS for the FIR absorption-line dominated ULIRG Mrk 231 by Fischer et al. (2010), is suggestive of a small rise in this ratio at this end in panel g. We find that the [NIII]57/[NII]122 ratio decreases linearly with decreasing ([OIII]+[NIII])/FIR, shown in panel h, with the exception of an outlier, Arp 299, while neither line is detected with LWS in Arp 220. The [OIII]52/[OIII]88 ratio distribution across the sequence, although also not measured for Arp 220, is best fit with a constant ratio, near unity, which is just above the low density limit of about 0.6 marked by a dashed line in panel i, and indicative of densities around $n_e = 10$ cm$^{-3}$ (Rubin et al. 1993). Similarly, we find little if any significant decrease in the PAH6.2/[CII]158 ratio with decreasing ([OIII]+[NIII])/FIR in panel j. Lastly, at the emission line end of the sequence, in the galaxies in which weak molecular absorption is detected, i.e. in NGC 2146 and M82, only absorptions from the ground state were detected (Table 3). In contrast, at the absorption-dominated end of the sequence, i.e.

in NGC 253, NGC 4945, and Arp 220, the highest excitation absorption extends to higher $E_{lower}$ with decreasing ([OIII]+[NIII])/FIR: 61 K for NGC 253 and NGC 4945, and 416 K for Arp 220.

In the next sections, we attempt to understand and quantitatively address the significant trends that accompany decreasing ([OIII]+[NIII])/FIR in our sample by comparing them to detailed spectral simulation models.

## 5. MODEL DESCRIPTION

The calculations presented in this work use version c13.00 of the spectral synthesis code *Cloudy* (Ferland et al. 2013), publicly available at www.nublado.org. Ferland et al. (1998) describe the treatment of the major processes that affect the ionization structure, while recent updates to the grain and molecular processes have been described by van Hoof et al. (2004), Shaw et al. (2005), Abel et al. (2005), and Paper I (and references therein). For a given set of input parameters, *Cloudy* computes the chemical and thermal structure of a cloud starting at the illuminated face of a hot, ionized $H^+$ region and continuing deep into regions of high column density and $A_V$ where most atoms have combined to form molecules. Relevant for simulation of interstellar medium (ISM) conditions in infrared-bright and infrared luminous galaxies, in addition to photo-ionization and photo-dissociation processes, *Cloudy* incorporates photoelectric heating of gas by grains as a function of grain properties such as temperature, charge, material and size distribution, a large number of molecules and reactions, molecule formation and condensation onto grains, and cosmic ray ionization and heating (van Hoof et al. 2004, Abel et al. 2005, Shaw et al. 2005, 2008, Ferland et al. 2013, and references therein).

The calculations presented here largely follow the same set of source spectral energy distributions (SEDs), model parameters and conditions given in Paper I, in which measured line and broadband ratios were predicted for both AGN and starburst central illumination in a spherical 1D geometry as a function of ionization parameter for cloud face density $n(H^+) = 1000$ cm$^{-3}$ and $A_V = 100$. In this work the model parameter set is modified and expanded to include 3 cloud face hydrogen densities $n(H^+) = 30, 300,$ and $3000$ and hydrogen column densities up to $N(H) = 10^{25}$ cm$^{-2}$, while the magnetic field at the face, $B_0$, was lowered compared with Paper I from 300 to 100 μG to be more compatible with current estimates of the magnetic field in the Orion HII region (Rao et al. 1998, O'Dell 2001). Here, as in Paper I, the nominal single AGN and starburst SEDs are fixed. In this sense, for simplicity we inherently assume that the heating is dominated by the youngest starbursts or an AGN, rather than by a combination of both. The relationship of the central source to the gas and dust in the galaxy is characterized by the ionization parameter $U$, defined as the ratio of hydrogen ionizing photon density to hydrogen density and equal to $\phi_H / n(H^+)c$, where $\phi_H$ is the flux of hydrogen ionizing photons, $n(H^+)$ is the hydrogen density at the illuminated face and $c$ is the speed of light. $U$ is varied from $10^{-4}$ to $10^0$ in increments of 1 dex, by varying the ionizing photon flux. We assume constant pressure throughout, which changes the hydrogen density with depth into the cloud, with no turbulent pressure other than a small turbulent broadening of 1 km s$^{-1}$ for its effects of line optical depths. In all calculations, at high $A_V$ magnetic pressure dominates and the density reaches a constant value. In the H$^+$ region, thermal and radiation pressure dominate, with radiation pressure dominating for higher $U$ models (>$10^{-2}$), see Abel et al. (2008) and Pellegrini et al. (2007). Table A1 in the Appendix summarizes the input parameters including the AGN and starburst SEDs, the range of ionization parameters, the cosmic ray ionization rate, the range of densities, $n(H^+)$, the equation of state and magnetic field dependence on $n(H^+)$, the gas phase abundances, dust and PAH properties, geometry and range of stopping column densities $N(H)$, i.e. the radius at which the calculation is stopped.

The model results are presented as contour plots in Figures 3–7. The results for the starburst SED are shown in the top row of each figure and those for the AGN SED are shown in the bottom row. Figures 3a–c present the results for the ([OIII]+[NIII])/FIR, [OIII]88/FIR, and [OIII]57/FIR ratios, Figures 4a,b present the results for $N(OH)$ and $N(H_2O)$, Figure 5 presents the results for F60/F100, Figures 6a-d present the present the results for the [CII]158/FIR, [NII]122/FIR [OI]145/FIR, and ([CII]+[OI])/FIR ratios, and Figures 7a-d present the results for the line ratios [OI]63/[OI]145, [OI]145/[CII], [NIII]57/[NII]122, and [OIII]52/[OIII]88, respectively. For each ratio or column density plot, the observed ratio ranges or derived $N(OH)$, $N(H_2O)$ ranges from the literature are indicated with shaded areas and arrows denoting upper/lower limits.

# 6. DISCUSSION

As discussed in section 4, with the decline of ([OIII]+[NIII])/FIR, the sample galaxies show the following trends: an increase in absorption line equivalent widths of OH, $H_2O$, and other molecules, in this study traced by the strong OH ground state transition at 119 μm, decreasing and then increasing F60/F100, and decreasing F25/F60, ([CII]+[OI])/FIR, [OI]145/FIR, [OI]63/[OI]145, and [NIII]57/[NII]122. The trends and values of these ratios provide clues to the origin of this dramatic sequence in the far-infrared spectra of galaxies, which we discuss in this section by comparing our *Cloudy* models of various line and total luminosity ratios with the observed ones.

The *Cloudy* models are of course idealizations of the real conditions that are present in galaxies. They assume that the source(s) of radiation are concentrated at the center of a surrounding spherically distributed cloud of gas and dust and can be represented by either the SED of a chosen young starburst or power-law AGN. Inhomogeneities, asymmetries, and combinations and distributions of sources cannot be taken into account. Moreover, specific starburst and AGN SEDs are assumed. For these reasons, although we compare the values of the observed fluxes and ratios with the models, we attach particular importance to trends in the models that agree with those observed along the sequence of decreasing ([OIII]+[NIII]) / FIR. In our synthesis of the results based on the totality of the diagnostics, we will discuss the possible effects of strong emission components that are associated with minor components in terms of their luminosity, may affect the various diagnostics when the line emission of more luminous components is in deficit.

## 6.1. Comparison of sample trends with models

### 6.1.1. ([OIII]+[NIII]) / FIR

In section 4 we described some of the trends observed with ([OIII]+[NIII])/FIR as it decreases in the sample from $3.51 \times 10^{-3}$ to less than $1.2 \times 10^{-4}$ for Arp 220, in which the [OIII] and [NIII] lines were not detected. Qualitatively, Figure 3a shows that for both starburst and AGN models a strong decrease in ([OIII]+[NIII])/FIR occurs with increasing N(H) and to a lesser extent with increasing density. This ratio, as well as the individual line ratios [OIII]88/FIR (Figure 3b) and [NIII]57/FIR (Figure 3c), also decrease at the high and low extremes of the ionization parameter for AGN in particular, but the trends are significantly weaker and the differences between AGN and starbursts are also relatively small over much of the relevant regions of parameter space.

Comparing the observations with these models more quantitatively, at the strong emission line end of the sequence, i.e. the bottom of the sequence in Figure 1, the value of $3.51 \times 10^{-3}$ indicates an upper limit on log $N(H)$ of about 22.5 for both starbursts and AGN, for all values of log $U$ and $n(H^+)$ in our explored parameter space. At the line deficit end of the sequence, this ratio is $< 10^{-4}$ for Arp 220 in which none of these lines were detected, indicating high $N(H)$. For example, for an ionization parameter as high as log $U = -1.5$ for both starburst and AGN models in Figure 3a, this limit is indicative of high hydrogen column densities $N(H) > 10^{24.2}$, $10^{24.1}$, and $10^{23.7}$ for $n(H^+) = 30$, 300, and 3000, respectively, although we note that for AGN models with even higher values of the ionization parameter or extremely low ionization parameters, the lower limits on the column density are less stringent. We note that because of the orders of magnitude predicted range in this ratio, the exact observed value in line deficit sources, can be significantly increased by energetically unimportant, but unobscured regions of star formation compared with the most luminous components whose line emission is in deficit due to FIR extinction and/or high density/high $U$ source(s). We note that line flux ratios of two lines in deficit are even more sensitive to this effect than line flux-to-FIR ratios because both numerator and denominator can be dominated by flux from an energetically unimportant region. An example of this could be a low luminosity region of star formation, possibly located in the outskirts of the nuclear region, with low $N(H)$, low $U$, and/or low $n(H^+)$.

*6.1.2. OH and $H_2O$ absorption column densities*

The nuclear OH and $H_2O$ column densities and their ratios can be compared to our *Cloudy* model predictions shown in Figures 4a, 4b, and 4c, respectively. The starburst models show a transition from UV-dominated chemistry to cosmic-ray-dominated chemistry at around column densities of $\sim 10^{21.5}$ cm$^{-2}$, corresponding to the H/H$_2$ and C$^+$/C/CO transition zones. At higher column densities, the OH to H$_2$O ratios increase with decreasing $U$. This is due to preferential condensation of $H_2O$ onto grain surfaces at the lower temperatures corresponding to the lower $U$ models. For the higher column density range all starburst models are driven by cosmic-ray ion-molecule chemistry. For the AGN models, deep penetrating X-rays keep the gas warm enough to prevent $H_2O$ condensation. In addition, the increased level of ionization lead to enhancement of $H_3O^+$, which leads to increased OH and $H_2O$ production. The $H_3O^+$ branching ratio is weighted towards OH production, leading to the enhanced OH to $H_2O$ column density ratios present in the AGN models.

We first discuss the molecular absorption side of the sequence (Arp 220, at the top end of Figure 1). González-Alfonso et al. (2004, 2012) derived $H_2O$ column densities toward the nuclear region component of Arp 220 of $\geq 0.8 - 6.0 \times 10^{18}$ cm$^{-2}$ per unit dust opacity at 50 microns, with an OH/$H_2O$ ratio between $0.4 - 1.0$ (see González-Alfonso et al. 2012). Based on analysis of multiple far-infrared absorption (and one emission) line strengths, these values are lower limits to the total column densities because they pertain only to the $\tau(50$ $\mu$m$) \sim 1$ outer surface of a galaxy whose dust continuum is thought to be optically thick even at 200 microns. Lower column densities and higher OH/$H_2O$ ratios were derived for the cooler, more extended components. Both the starburst and AGN models produce a strong increase in OH and $H_2O$ column densities with increasing $N(H)$, and with increasing ionization parameter for $N(H) \geq 10^{22.5}$ cm$^{-2}$, for intermediate and high densities. The AGN models generally produce higher OH/$H_2O$ ratios. For the starburst models, only the lowest density $n(H^+) = 30$ cm$^{-3}$ models at the very highest values of $N(H)$ and $U$, are able to produce the high column densities $N(OH)$ and $N(H_2O)$ and with the ratio derived by González-Alfonso et al. (2012). The higher density $n(H^+) = 300, 3000$ cm$^{-3}$ starburst models, which likely more closely reflect the conditions in ULIRGs, underpredict the observed OH column densities and/or the OH/$H_2O$ ratios over the parameter space that we have explored. In contrast, the AGN models produce higher OH column densities than the starburst models, typically by an order of magnitude for the higher $n(H^+)$ models, and are in better agreement with what is observed in Arp 220. The intermediate and high density AGN models predict OH/$H_2O \sim 1$, consistent with the derived ratio estimate in Arp 220, at very high $U$ and $N(H)$. For example, for our highest density models, at log $U = -0.6$ and log $N(H) \geq 24.8$, OH/$H_2O \sim 1$ and, as we discuss in the next sections, the model F60/F100 ratio is close to the measured ratio for Arp 220. We note also that for the highest values of $N(H)$, it is the highest values of $U$ that produce the very high OH and $H_2O$ column densities in the outer layers, just the regions probed by the FIR molecular profiles when the dust emission is optically thick.

In the starburst galaxies NGC 2146, Arp 299[7], and M82 at the strong emission line side of the sequence, the equivalent widths of the ground state OH119 absorption features are similar and relatively low: 0.026, 0,029, and 0.021 microns, respectively (Table 2, column 13), and no excited OH lines were detected with LWS, suggesting low column densities. Based on the equivalent width in M82, upper limits on transitions from excited levels, and the assumptions that the line is optically thin and predominantly in the ground state, Colbert et al. (1999) estimated the OH column density to be $\sim 4 \times 10^{14}$ cm$^{-2}$. Although $H_2O$ is not detected in the LWS spectra of the galaxies at the strong emission line end of the sequence, in M82, Herschel HIFI and SPIRE observations of absorption from the ground state in the para-$H_2O$ ($1_{11} - 0_{00}$) line indicate column densities in the range $4 - 9 \times 10^{13}$ cm$^{-2}$ (Weiβ et al. 2010; Kamenetzky et al. 2012), a factor of $5 - 10$ times lower than the OH column density estimated from the OH119 feature in M82. Based on our starburst models and M82's high value of ([OIII]+[NIII])/FIR (log ([OIII]+[NIII])/FIR = -2.58), these absolute OH and $H_2O$ column densities and OH/$H_2O$ ratios are most consistent with the low $N(H)$, low ionization parameter regime (log $N(H) \sim 21.3$, log $U \sim$ -3.3), and a face density of $n(H^+)$=3000 (Figures 4a, b). Our AGN models, on the other hand, produce at least a factor of $\sim 3$ higher OH/$H_2O$ ratios than are observed, at M82's observed ([OIII]+[NIII])/FIR values and OH column density.

### 6.1.3. F60/F100

Figure 5 presents the F60/F100 ratios predicted for the range of parameters in our *Cloudy* models and the grey shaded region highlights the range of values observed in the sequence. These models indicate that for a given density $n(H^+)$, this ratio range defines relatively narrow, "diagonal" bands in log $U$, $A_v$ parameter space. In general, the F60/F100 colors of the sequence galaxies can be due to a combination of low far-infrared extinction and low $U$ or to higher extinction in combination with higher U. The F25/F60 ratio displayed in Figure 2c remains fairly flat over most of the sequence, with the exception of the outlier NGC 1068 as mentioned in

---

[7] The Arp 299 interacting/merging system has complex structure containing several nuclei (e.g. Gehrz et al. 1983, Aalto et al. 1997, Satyapal et al. 1999). The full LWS spectrum, which includes both nuclear and diffuse emission components, is emission-line dominated, may mask the characteristics of the individual components.

Section 4, but increases for the two galaxies at the molecular-absorption dominated end, also suggesting higher $N(H)$ at this end of the sequence. The range of the F60/F100 ratio in our sample extends from 0.46 for IC 342 to 1.08 for M 82, while for the relatively "cool" ULIRG Arp 220 the ratio is 0.90 (Figure 2b and Table 1, column 7) and for the 21 ULIRGs listed in the IRAS RBGS the average ratio is 0.98. For Arp 220, our *Cloudy* AGN models with $n(H^+)$ = 3000, log $U$ = - 0.6, log $N(H) \geq 24.8$ best reproduce the observed F60/F100 ratio and high $N(H)$, $N(OH)$, and $N(H_2O)$ column densities derived from fits to the Herschel PACS Arp 220 spectrum by González-Alfonso et al. (2012). These model parameters are consistent with the upper limits on ([OIII]+[NIII])/FIR, but produce OH/$H_2O$ ~ 2, which is a factor of 2-5 higher than the range derived by González-Alfonso et al. (2012). The model trend with density suggests that increasing the density above $n(H^+)$ = 3000, our highest density model, could result in a decrease in the OH/$H_2O$ ratio to within the range derived from observations.

At the high emission line side of the sequence, M82's high F60/F100 ratio of 1.08, is consistent with the starburst parameters derived in the previous section (log $U$ = -3.3, $n(H^+)$ = 3000 cm$^{-3}$, and log $N(H)$ = 21.3), based on the values of ([OIII]+[NIII])/FIR, the OH and $H_2O$ column densities, and the OH/$H_2O$ ratios.

For NGC 1068, if we assume, very approximately, that ionization and heating from the AGN and starburst ring contribute approximately equally to both the ([OIII]+[NIII]) and FIR emission (Telesco et al. 1984), then both can be characterized by the ratio measured in the LWS beam, ([OIII]+[NIII])/FIR = 2.17 x 10$^{-3}$. While we find that moderate density starburst models with high ionization parameters, can be found that are consistent with this value and with the measured F60/F100 value of 0.763 (log F60/F100 = -0.12), they do not produce the high measured [CII]158/FIR value of 1.6 x 10$^{-3}$ (log [CII]158/FIR = -2.8, Figures 3a, 5, and 6a). In contrast, a high density ($n(H^+)$=3000 cm$^{-3}$), low ionization parameter (log $U$=-3.3), low column density (log $N(H)$ cm$^{-2}$ = 21.5) starburst model similar to that chosen for M82 fits all of these measured values. This is the model we adopt for the starburst component of NGC 1068. We do not attempt here to estimate the conditions in the AGN dominated ionized regions in NGC 1068, as the LWS observations are hampered by relatively low spatial resolution and contamination by the extended starburst.

*6.1.4. [CII]158/FIR, ([CII]+[OI])/FIR, [NII]122/FIR*

The [CII]/FIR deficit for the sample galaxies along the sequence is evident from Figure 1 and was the focus of Paper I, Luhman et al. (2003), Malhotra et al. (2001) and numerous observational studies. As pointed out in Helou et al. (2001), a plausible explanation could be that due to high densities, the gas cooling is dominated by the [OI]63,145 lines for the deficit galaxies. As is evident from Figure 1, these lines as well as the [NII]122 line, also decrease in flux relative to the FIR flux at this molecular absorption end of the sequence, so this explanation is ruled out as Helou et al. also concluded. Figures 6a,b show the *Cloudy* model results individually for the [CII] and [NII] lines which show very similar trends in the models, dropping by orders of magnitude for both high $U$ and high $N$(H), and also decrease with increasing density, while the [OI]145/FIR and ([CII]+[OI])/FIR ratios (Figures 6c,d, respectively) are much less sensitive to density. Figure 2d shows that the sum of the measured PDR cooling lines of $C^+$ and atomic oxygen relative to the FIR luminosity is fairly constant at a level of about $3.5 \times 10^{-3}$ (log ([CII]+[OI])/FIR = -2.46) over most of the sequence before it drops for the three galaxies with the highest molecular absorption. For starburst models, when log ([CII]+[OI])/FIR = -2.46, log $N$(H) is always less than 23.5 and log $U$ always less than -2.5 for all three values of $n(H^+)$. This is broadly consistent with the conclusions drawn about galaxies on the bright emission line side of the sequence in the earlier sections based on their ratios of (OIII+NIII)/FIR, F60/F100, and estimated column densities of OH and $H_2O$. However, for M 82, for which the $n(H^+) = 3000$ models were favored in earlier sections, lower densities are preferred for the ([CII]+[OI])/FIR ratio, primarily because of the relatively high [OI]145/FIR ratios predicted by our models, as we discuss in detail in the next section. In Arp 220, the value of about $3.7 \times 10^{-5}$ is observed for this ratio with LWS, but we note that only an upper limit on the [OI]145 line was observed with LWS while absorption in the [OI]63 line contributed negatively to the ratio. In AGN models, preferred for Arp 220 in this work in order to produce the high observed column densities of OH without producing too much $H_2O$, the conditions suggested in the previous section, $n(H^+) \geq 3000$, log $U$ = - 0.5, log $N$(H) $\geq$ 24.8, are consistent with the observed values of this ratio.

### 6.1.5. [OI]145/FIR

The average [OI]145/FIR ratio for the seven galaxies on the high emission line end of the sequence is 1.1 x $10^{-4}$, decreasing to an upper limit in Arp 220 of < 2 x $10^{-5}$ on the line deficit end of the sequence, as is plotted in Figure 2e. In contrast with other far-infrared fine structure line flux to FIR ratios, this ratio increases with density over almost our entire range of parameter space (Figure 6c), owing to the high critical density of the [OI]145 line. This makes it a very powerful diagnostic for examining whether high density is primarily responsible for the fine-structure line deficits. Since the measured [OI]145/FIR ratios decrease at the line deficit end of the sequence in Figure 2e, we conclude that high density is not the dominant cause of the line deficits at the top of the sequence in Figure 1 (see also Fischer et al. 2010 for Mrk 231). The ratio of ~ $10^{-4}$ for most of the sequence is consistent with our finding of relatively low $N$(H) and low $U$, but only for the case of low density in contrast to what is suggested by the diagnostics discussed so far in M82 and NGC 1068 (([OIII]+[NIII])/FIR, $N$(OH), $N$($H_2O$), and F60/F100) for which the high density case is preferred. Ratios of < 2 x $10^{-5}$, as found for Arp 220, are indicative of high $U$ and/or high $N$(H) for both starburst and AGN models. Here again, we note that, particularly for line fluxes that are in deficit, emission from energetically unimportant regions can dominate the emission from regions with deficits, and thus line flux-to-FIR ratios should be regarded as indicative of conditions, but not accurate diagnostics of conditions.

### 6.1.6. [OI]63/[OI]145

The measured [OI]63/[OI]145 ratio is approximately constant in Figure 2f for the majority of the galaxies in the sequence, showing a slight peak in the middle of the sequence increasing from ~ 13.5 to ~ 20 for the AGN galaxies NGC 1068 and Cen A and then decreasing again for line deficit galaxies dropping to ~ 8 for NGC 4945. The very similar starburst and AGN *Cloudy* model results for all three densities (Figure 7a), showing vertical contours in log $U$ versus log $N$(H), provide strong support for low values of $N$(H) on the emission line side of the sequence with increasing values of $N$(H) toward the line deficit end of the sequence with log $N$(H) as high as 23.5 for NGC 4945. The somewhat higher ratios found in NGC 1068 and Cen A are consistent with somewhat higher density and/or ionization parameter regions, also at low $N$(H).

### 6.1.7. [OI]145/[CII]

The [OI]145/[CII] ratio in the sample galaxies does not show much of a trend in Figure 2g, and the values for all galaxies are within a factor of three of each other. M82 shows the highest value of this ratio (~ 0.11) and the lowest values (~ 0.035) appear in the mid-range of the sequence, with a hint of a rise again toward the molecular absorption end (~ 0.08). Because of the much higher critical density of the [OI]145 line compared with that of [CII]158, the [OI]145/[CII]158 ratio is sensitive to density and in our *Cloudy* models it rises by an order of magnitude to values of order unity at the highest densities for both starburst and AGN models (Figure 7b). In addition it is sensitive to the ionization parameter in our models (see discussion in Paper I). In contrast, because of the relatively long and similar wavelengths of the [OI]145 and [CII]158 lines, the ratio is very insensitive to $N$(H). The slight increase in the ratio to 0.08 on the line deficit end of the sequence could be due to higher densities or higher ionization parameters, however none of the galaxies show measured values near the ~ unity values of the starburst and AGN models at our highest density grid point $n(H^+) = 3000$. On the other hand, for the molecular absorption line end of the sequence, the ratio of 0.08 is consistent with moderate densities and high column densities of log $N$(H) ≥ 24.5 and low to moderate ionization parameters log $U$ ≤ -1.5, while the mid-sequence ratio of 0.035 is consistent with low densities for starburst models and low $U$ and a range of column densities for AGN models.

### 6.1.8. [NIII]57 / [NII]122 ratio

The measured [NIII]57/[NII]122 ratio in Figure 2h shows an almost linear trend with ([OIII]+[NIII])/FIR, varying from about 2 at the emission line end to 0.25 at the absorption line end, with one outlier, Arp 299 for which the measured ratio is about 5. Density enhancement at the deficit end of the sequence would have the opposite effect to that observed, since the critical density of [NIII]57 is nearly an order of magnitude higher than that of [NII]122. The lowest value of the ratio, 0.25, measured for NGC 4945 is indicative of high $N$(H), generally greater than $10^{24.0}$ for both starburst and AGN models for all three densities (Figure 7c), although lower column densities are predicted for the lowest ionization parameters. At the emission line end of the sequence, the ratio of ~ 2 is consistent with low $N$(H) and low ionization parameter for the highest density for starbursts and for all densities for AGN models. We note that this trend is consistent with the finding of Zhao et al. (2013), using both ISO/LWS [NII]122 measurements and Herschel/SPIRE [NII]205 measurements, that these line luminosities are linearly correlated with $L_{IR}$ for $L_{IR} < 10^{12}\,L_\odot$, while a deficit is found in [NII]205/$L_{IR}$ for $L_{IR} > 10^{12}\,L_\odot$.

*6.1.9. [OIII]52 / [OIII]88 line ratio*

The measured [OIII]52/[OIII]88 line ratio shows large scatter around unity, no trend with ([OIII]+[NIII])/FIR and is consistent with densities at or above the low density value of ~ 0.6 for 9 sequence galaxies (Figure 2i). The ratio is undetermined for Arp 220, since neither line was detected. This ratio, nearly identical for our starburst and AGN models (Figure 7d), is very sensitive to column density at high column density and sensitive to density and ionization parameter at low column densities. The measured values are consistent with both high and low ionization parameters for moderate densities and for the highest densities with column densities ~ 1 x $10^{-23.75}$ cm$^{-2}$. The lack of a trend for the [OIII]52/[OIII]88 ratio may indicate that the energetically important line deficit regions of galaxies are often overwhelmed by energetically unimportant regions with strong line emission.

*6.1.10. PAH-6.2 / [CII]158*

The PAH-6.2/[CII]158 ratio appears relatively flat over the entire range of the sequence, perhaps with a slightly downward trend toward the line deficit end (Figure 2j). This result is similar to the finding of Helou et al. (2001) for their sample of normal galaxies, that the
(5 − 10 μm)/[CII]158 ratio is constant while the [CII]158/FIR and (5 − 10 μm)/FIR ratios drop with increasing F60/F100 ratio, where the broadband 5 − 10 μm flux was taken as a measure of the PAH feature emission, and the finding of Luhman et al. (2003) who found a relatively constant value of the PAH-6.2/[CII]158 ratio for their larger sample of both normal galaxies and ULIRGs. The lack of a significant decreasing trend in this ratio, one involving two lines with over an order of magnitude factor in wavelength, suggests again that the ratio may be dominated by the line emission in regions of the galaxy not affected by severe extinction by dust, i.e. regions external to the FIR luminous, dust-obscured regions, consistent with the results of Díaz-Santos et al. (2011, 2014) that both PAH emission and [CII]158 line emission are commonly extended in LIRGs, that the PAH emission is more extended than both the mid-IR continuum and the [NeII]12.8 line emission, and that only the nuclear components of LIRGs show [CII]/FIR line deficits. This conclusion is justified in particular for galaxies like the two highest line deficit galaxies in the sample, Arp 220 and NGC 4945, for which very high obscuring column densities have been inferred (e.g. González-Alfonso et al. 2012, Marinucci et al. 2012). So again we suggest that this is due to the fact that, especially for features that are in deficit, energetically unimportant, emission from regions *not* in deficit can dominate line flux-to-FIR ratios and line-to-line flux ratios. The latter are most sensitive to this effect, when both quantities are in deficit. We thus consider results based on line-to-FIR ratios and on integrated properties such as molecular column densities, and broadband FIR color ratios, more indicative of the dominant global conditions than those based on ratios of lines in deficit. Ratios of such line-to-FIR flux can provide lower limits to global conditions rather than absolute values, for the same reason.

## 6.2. Trends summary and conclusions

Here we summarize the observational trends that accompany decreasing ([OIII]+[NIII])/FIR in our sample of galaxies and the trends in physical conditions suggested by our models. With decreasing ([OIII]+[NIII])/FIR, we find generally increasing OH119 absorption, accompanied by decreasing and then increasing F60/F100, and decreasing F25/F60, ([CII]+[OI])/FIR, [OI]145/FIR, [OI]63/[OI]145, and [NIII]57/[NII]122. According to our idealized models of AGN and starbursts, these observed trends in the sequence of decreasing atomic and ionic fine-structure line fluxes together with increasing equivalent widths of OH, $H_2O$ and other molecules, require both increasing column densities *and* increasing ionization parameters, while density appears to play less of a role. Although increasing density is also consistent with some of the spectral trends in the sequence, the decreasing trend in [OI]145/FIR is opposite to what is predicted if increasing density were primarily responsible for the spectral trends in the sequence.

In our models, high ISM column densities, on the order of $N(H) \sim 10^{24}$ cm$^{-2}$ and greater are consistent with i) the high column densities of OH and H$_2$O observed in the nearest ULIRGs (González-Alfonso et al. 2012, 2014), ii) extinction of far-IR emission lines due to dust, including atomic and ionized fine-structure lines and molecular emission features such as the often faint redshifted emission in P-Cygni profiles of OH (Fischer et al. 2010, Sturm et al. 2011, Veilleux et al. 2013, Spoon et al. 2013, González-Alfonso et al. 2014), producing positive total equivalent widths in the OH119 ground-state line in combination with the OH blueshifted absorption components of molecular outflows, and iii) depressed values of mid-IR line and continuum fluxes relative to far-IR broadband fluxes. In their study of the velocity profiles of the OH119 and OH79 ground-state lines used to trace molecular outflows in ULIRGs, Spoon et al. (2013) and Veilleux et al. (2013) point out that the total equivalent widths of the OH119 profiles, including both absorption and emission components, are correlated with the mid-infrared silicate absorption feature strengths. This can be understood, as Spoon et al. (2013) point out, and as is discussed in detail by González-Alfonso et al. (2014), because for the case of pure radiative excitation, i.e. negligible collisional excitation of the OH119 doublet, and in spherical symmetry, the total OH119 equivalent width integrated over all velocities is expected to be zero unless there is significant far-infrared continuum optical depth, and we note that it is this total OH119 equivalent width that is measured with LWS, owing to its low velocity resolution. Further, as mentioned earlier, in their study of fine-structure lines in ULIRGs, Farrah et al. (2013) find that for ULIRGs above a threshold silicate strength, there is a correlation between silicate strength and the severity of the [CII]158 and [NII]122 line deficits, consistent with our results.

High $U$ is needed in the molecular-absorption dominated galaxies to maintain the relatively warm far-IR colors of molecular-absorption dominated galaxies such as ULIRGs, despite the high $N(H)$ and associated far-IR reddening. Increasing ionization parameters at the molecular-absorption end of the sequence are also consistent with the increase in excitation of the molecular absorption lines. For ionization parameters above $10^{-2}$, the UV photon density per hydrogen density is so high at the face of the cloud that the UV photons are preferentially absorbed by dust (Voit 1992; Bottorff et al. 1998), producing high far-infrared radiation density further into the cloud where the OH and H$_2$O molecules are pumped to higher levels by the abundant far-infrared photons (González-Alfonso et al. 2004, 2008, 2010, 2012).

Our models suggest that $N$(H) varies from values as low as $10^{21}$ cm$^{-2}$ to values as high as $10^{24.8}$ cm$^{-2}$ and greater, depending on the exact derived molecular column densities which, at the absorption line end of the sequence, can only be determined for the regions outside of unity FIR optical depth. In addition, log $U$ increases from values as low as -4.0, to values as high as -1.5 to 0.0. For Arp 220, the best overall matches are with the $n$(H$^+$) = 300 – 3000 cm$^{-3}$ AGN models. In contrast, the high emission line, low absorption end of the sequence appears to be best explained with starburst models with low $N$(H) and low $U$. At the absorption-dominated end of the sequence, the high density case better reproduces the OH column densities derived from Herschel data for Arp 220 (González-Alfonso et al. 2012), while the moderate density case better matches the ionized line deficits and line ratios. Other ratios, such as the fine-structure line-to-FIR ratios are consistent with these inferred global parameters. However we find that line ratios such as [O I]145/[C II]158, [O III]52/[O III]88, and PAH-6.2/[C II]158 do not yield consistent results, which we attribute to their stronger sensitivity to the effects of foreground emission by non-nuclear, energetically less significant star forming regions that are not included in the models, which treat only a single source of radiation surrounded by a spherically symmetric, smoothly distributed medium.

We emphasize the importance of both high far-infrared continuum optical depth and high ionization parameters and the associated high far-infrared photon densities, for understanding mid- and far-infrared fine-structure line diagnostics as well as the structure and power sources of the feedback traced by molecular outflows. Indeed high far-infrared continuum optical depth can in principle hide both velocity profile signatures of molecular outflows if the outflows have preferential bipolar directions and AGN signatures if the obscuring dust is both optically and physically thick (see also Spoon et al. 2013, Veilleux et al. 2013).

Lastly and importantly, we note that our models are based on a number of assumptions about the starburst and AGN SEDs, magnetic field levels, equation of state, and cosmic ray ionization rates that may not be representative of some or all of the galaxies in the sample. In particular, while we use a constant cosmic ray ionization rate of 5 x $10^{-17}$ s$^{-1}$, as in Abel et al. (2005, see their Appendix C for details), we note that there is recent evidence for the presence and importance of high cosmic ray densities and ionization rates due to supernovae and massive stellar winds in star forming galaxies (VERITAS Collaboration et al. 2009; Acero et al. 2009; Bayet et al. 2011; Lacki et al. 2011; Meijerink et al. 2011; Papadopoulos et al. 2011) with significant effects on chemistry of molecules such as OH and H$_2$O and related molecular ions (González-Alfonso et al. 2013). In this work, as a first step, we have kept the cosmic ray ionization rate constant while varying $n(H^+)$, $N(H)$, and $U$ for both AGN and starburst model SEDs.

## 7. SUMMARY AND FUTURE WORK

In this work we have extended our earlier work using the *Cloudy* spectral synthesis code for central starburst and AGN by varying both $N(H)$ and $n(H^+)$ in addition to $U$, in order to identify the physical changes primarily responsible for the dramatic sequential progression observed in IR-bright galaxies, extending from strong fine-structure line emission to detection of only faint [CII]158 μm fine-structure line emission, accompanied by strong molecular absorption. The molecular absorption spectra show varying excitation as well, extending from galaxies in which the molecular population mainly occupies the ground state to the nearby ultraluminous galaxy Arp 220 in which there is significant population in higher levels. Our analysis is based primarily on data taken with the ISO LWS but also incorporates published OH and H$_2$O column densities derived from ISO and Herschel Space Observatory spectroscopy. We have compared trends in line-to-total far-infrared ratios, IRAS colors, and OH and H$_2$O column densities with a grid of models produced with the latest version of the *Cloudy* spectral synthesis code, in which we logarithmically vary the hydrogen density at the face of the cloud $n(H^+)$, the ionization parameter $U$, and the hydrogen column density of the cloud $N(H)$, for central starburst and AGN SEDs, with an isobaric equation of state including pressure due to thermal, radiation, and magnetic field and a constant cosmic ray ionization rate of 5 x 10$^{-17}$ s$^{-1}$, as was used in Paper I. We find that the sequence from strong emission to strong absorption can best be explained by increasing both the hydrogen column density, from as low as 10$^{21}$ cm$^{-2}$ to as high as 10$^{24.8}$ cm$^{-2}$ or greater, and the ionization parameter, from as low as 10$^{-4}$ to as high as 1, for our n(H$^+$) = 300 – 3000 cm$^{-3}$ models. The starburst models best explain most of the sequence, while the AGN models are better able to explain the high OH and H$_2$O column densities in Arp 220. We note that high cosmic ray fluxes associated with high supernovae explosion rates, which we have not taken into account in the present models, could also mimic these effects in Arp 220.

The result that the sequence transitions to high ionization parameter at the absorption line end of the sequence is also consistent with the changes in the character of the molecular absorption spectrum. At the emission line end of the sequence, in the galaxies in which weak molecular absorption is detected, i.e. in NGC 2146 and M82, only ground state transitions were detected. In contrast, at the absorption line end of the sequence, i.e. in NGC 253, NGC 4945, and Arp 220, the highest excitation absorption extends to higher $E_{lower}$ with decreasing ([OIII]+[NIII])/FIR. Our results suggest that the molecular ISM in ULIRG-like, absorption dominated systems is located close to and at least partially covers the source of excitation with high molecular column densities. We note that the ISM geometry may be disk-like or torus-like, as is suggested for Mrk 231 (Klöckner, Baan, & Garrett 2003, González-Alfonso et al. 2014), probably formed as a result of dissipative collapse during the merger process, and appears to be feeding the massive molecular outflows found by Herschel studies to be ubiquitous in these systems. Such departures from spherical symmetry and viewing angle effects will need to be better characterized in order to further our understanding of the power sources and drivers of feedback in these systems.

The type of analysis we present in this work is needed for the interpretation of present and future far-infrared and submillimeter spectroscopic observations of (U)LIRGs by *Herschel*, ALMA, SPICA, CCAT and beyond, for the interpretation of both AGN and outflow diagnostics, as discussed in section 6.2. As mentioned in earlier sections, we have not yet incorporated variation of the low energy cosmic ray densities and ionization rates due to supernovae and massive stellar winds in star forming galaxies and from AGN. Incorporation of these effects and others, as well as prediction of the molecular line fluxes and profiles measured in Herschel studies, are important advances that we hope to address in future work by expanding the basic framework presented here.

We would like to thank Gary Ferland and the rest of the Cloudy group for useful discussions related to this work. Basic research in infrared astronomy at the Naval Research Laboratory is funded by the Office of Naval Research. E.G-A is a Research Associate at the Harvard-Smithsonian Center for Astrophysics, and thanks the Spanish Ministerio de Economía y Competitividad for support under projects AYA2010-21697-C05-0 and FIS2012-39162-C06-01. JF and E.G-A acknowledge support from NASA ADAP program NNX07AH49G. PvH acknowledges support from the Belgian Science Policy Office through the ESX PRODEX program.

## APPENDIX

The details of the model parameter set broadly described in Section 5, and used to produce the model results presented in Figures 3–7, are tabulated in Table A1.

## Table 1
### Far-Infrared Bright Spectroscopic Sample

| Galaxy | R.A. (J2000) | Decl. (J2000) | $cz$ (km s$^{-1}$) | $D$ (Mpc) | Aperture (kpc) | $F_{60}/F_{100}$ | log $L_{FIR}$ ($L_\odot$) | log $L_{IR}$ ($L_\odot$) | Type | Notes |
|---|---|---|---|---|---|---|---|---|---|---|
| ARP 220 | 15 34 57.19 | +23 30 11.4 | 5450 | 79.90 | 29.0 | 0.903 | 12.15 | 12.21 | Merger | ULIRG (nearest); Liner |
| NGC 4945 | 13 05 26.47 | -49 27 54.5 | 560 | 3.42 | 1.2 | 0.470 | 10.41 | 10.48 | SB(s)cd | Inclination=78°; hard X-ray source, Sy 2 |
| NGC 253 | 00 47 33.24 | -25 17 17.9 | 261 | 3.10 | 1.1 | 0.751 | 10.29 | 10.44 | SAB(s)c | Inclination=78°; |
| IC 342 | 03 46 49.71 | +68 05 44.8 | 31 | 4.60 | 1.7 | 0.462 | 10.01 | 10.17 | SAB(rs)cd | Face-on |
| M 83 | 13 37 00.30 | -29 51 51.2 | 522 | 3.60 | 1.3 | 0.507 | 9.94 | 10.10 | SAB(s)c | Face-on |
| CEN A | 13 25 27.58 | -43 01 08.4 | 562 | 4.03 | 1.5 | 0.518 | 9.94 | 10.11 | S0 pec | Elliptical+spiral, edge on, radio galaxy, BL Lac |
| NGC 1068 | 02 42 40.75 | -00 00 47.5 | 1005 | 13.70 | 5.0 | 0.763 | 10.89 | 11.27 | (R)SA(rs)b | Face-on; Sy 2 |
| M 82 | 09 55 52 21 | +69 40 46.8 | 187 | 3.63 | 1.3 | 1.078 | 10.61 | 10.77 | I0 | Irregular, disturbed morphology |
| ARP 299 | 11 28 32.42 | +58 33 44.0 | 3159 | 47.74 | 17.4 | 1.015 | 11.74 | 11.88 | Sbrst/? | Interacting pair |
| NGC 2146 | 06 18 38.77 | +78 21 23.8 | 885 | 16.47 | 6.0 | 0.756 | 10.93 | 11.07 | SB(s)ab pec | Disturbed morphology |

Notes. Entries in columns 2-5, 7-9 taken from the RBGS (Sanders et al. 2003).

**Table 2**

LINE FLUXES AND INTEGRATED CONTINUUM MEASUREMENTS

| Galaxy | FIR IRAS | FIR LWS | [CII] 158 | [OI] 145 | [NII] 122 | [OIII] 88 | [OI] 63 | [NIII] 57 | [OIII] 52 | PAH 6.2 | E.W.(OH) ($\mu$m) 119 | ([OIII]+[NIII])/FIR $10^{-3}$ |
|---|---|---|---|---|---|---|---|---|---|---|---|---|
| ARP 220 | 4.84 | 6.21 | 0.85 (0.04) | <0.13 | <0.12 | <0.63 | -0.62 (0.08) | <1.1 | <0.49 | 2.70 (0.20) | 0.134 (0.008) | <0.12 |
| NGC 4945 | 37.09 | 38.66 | 31.86 (0.43) | 2.20 (0.49) | 7.66 (0.60) | 6.44 (0.61) | 17.93 (0.54) | 1.92 (0.35) | 4.74 (1.0) | 64.30 (3.40) | 0.053 (0.003) | 0.34 |
| NGC 253 | 47.69 | 60.35 | 49.85 (0.40) | 4.26 (0.21) | 11.53 (0.65) | 7.60 (0.51) | 37.28 (0.77) | 4.37 (0.36) | 11.68 (1.3) | 174 (5.80) | 0.042 (0.002) | 0.39 |
| IC 342 | 10.81 | 6.54 | 15.16 (0.25) | 0.76 (0.09) | 3.38 (0.13) | 0.91 (0.32) | 7.58 (0.32) | 2.5 (0.36) | <3.0 | 43.00 (2.40) | <0.013 | 0.67* |
| M 83 | 15.25 | 8.51 | 17.54 (0.39) | 0.71 (0.08) | 3.43 (0.21) | 3.23 (0.44) | 11.77 (0.31) | 3.07 (0.30) | 3.62 (1.3) | 79.60 (3.90) | <0.009 | 1.17 |
| CEN A | 12.12 | 8.25 | 26.94 (0.53) | 0.97 (0.07) | 1.84 (0.12) | 7.66 (0.28) | 19.12 (0.51) | 1.40 (0.21) | 5.08 (0.89) | 51.70 (5.20) | <0.011 | 1.71 |
| NGC 1068 | 9.63 | 12.50 | 20.10 (0.22) | 0.69 (0.04) | 4.19 (0.14) | 10.33 (0.25) | 15.70 (0.32) | 5.28 (0.25) | 11.53 (0.34) | 67.20 (9.60) | -0.014 (0.002) | 2.17 |
| M 82 | 65.43 | 80.43 | 126.70 (1.2) | 13.84 (0.25) | 20.42 (0.66) | 81.79 (1.4) | 176.33 (4.1) | 30.51 (1.2) | 100.88 (3.3) | 1000.00 (18) | 0.021 (0.002) | 2.65 |
| ARP 299 | 5.08 | 6.16 | 7.56 (0.07) | 0.57 (0.02) | 0.35 (0.06) | 7.66 (0.31) | 8.29 (0.28) | 1.72 (0.19) | 7.42 (0.65) | 20.50 (3.10) | 0.029 (0.005) | 2.73 |
| NGC 2146 | 7.21 | 9.40 | 25.06 (0.07) | 1.36 (0.06) | 2.52 (0.16) | 15.19 (1.1) | 17.98 (0.80) | 5.40 (0.71) | 12.37 (1.1) | … | 0.026 (0.002) | 3.51 |

NOTES. Column 2 lists the FIR flux as estimated by the formula given in Sanders & Mirabel (1996), while column 3 lists the FIR flux integrated over the LWS range within the LWS aperture. Lines are denoted by their wavelengths in microns. FIR flux units are $10^{-12}$ W m$^{-2}$, line flux units are $10^{-15}$ W m$^{-2}$, and negative line fluxes indicate that the line is in absorption. Estimated rms errors are listed in parentheses and upper limits are 3$\sigma$ values. The equivalent width (E.W.) in microns is listed for the OH ground-state, $\Pi_{3/2}$ ladder, $\Lambda$-doublet at 119 $\mu$m. A negative equivalent width indicates the line is observed in emission. ([OIII]88+[OIII]52+[NIII]57)/FIR, is given in column 13 in units of $10^{-3}$ and is based on the far-infrared flux integrated over the LWS aperture listed in column 3. (*)For IC 342, the ([OIII]+[NIII])/FIR value is estimated based on the measured value of [OIII]88 and the median measured value of [OIII]52/[OIII]88 = 1.04±0.13.

OBSERVATION LWS TDT NUMBERS USED: NGC 253 (2470103, 56901708); NGC 1068 (60501183, 60500401); IC 342 (64600302, center); NGC 2146 (67900165); M 82 (18000501, 19400538); ARP 299 (18000704); NGC 4945 (28000440); CEN A (63400464); M 83 (44700575); ARP 220 (27800202, 64000801, 64000916).

**Table 3**

MOLECULAR LINE EQUIVALENT WIDTHS AND LOCAL CONTINUUM MEASUREMENTS

| MOLECULE/TRANSITION | $\lambda_0$ | $E_{lower}$(K) | ARP 220[2] | | NGC 4945 | | NGC 253 | | NGC 1068 | | M 82 | | NGC 2146 | |
|---|---|---|---|---|---|---|---|---|---|---|---|---|---|---|
| OH $\Pi_{1/2} - \Pi_{3/2}$ J = 3/2 – 3/2 | 53.3 | 0.07 | 116.4 | 3.7(0.2) | 477.6 | 1.9(0.2) | 1026.7 | 0.8(0.1) | | | | | | |
| $H_2O$ $4_{22} - 3_{13}$ | 57.6 | 204.71 | 110.3 | 0.8(0.1) | | | | | | | | | | |
| $H_2O$ $4_{32} - 3_{21}$ | 58.7 | 305.25 | 108.8 | 0.8(0.2) | | | | | | | | | | |
| OH $\Pi_{3/2} - \Pi_{3/2}$ J = 9/2 – 7/2 | 65.2 | 290.86 | 96.6 | 2.3(0.2) | | | | | | | | | | |
| $H_2O$ $3_{30} - 2_{21}$ | 66.4 | 95.53 | 94.3 | 1.7(0.1) | | | | | | | | | | |
| $H_2O$ $3_{31} - 2_{20}$ | 67.1 | 194.09 | 92.7 | 0.8(0.2) | | | | | | | | | | |
| OH $\Pi_{1/2} - \Pi_{1/2}$ J = 7/2 – 5/2 | 71.1 | 415.68 | 83.6 | 0.8(0.3) | | | | | | | | | | |
| $H_2O$ $3_{21} - 2_{12}$ | 75.4 | 114.38 | 75.8 | 1.7(0.2) | | | | | | | | | | |
| OH $\Pi_{1/2} - \Pi_{3/2}$ J = 1/2 – 3/2 | 79.1 | 0.04 | 70.3 | 2.3(0.2) | | | | | 137.6 | -1.0(0.2) | | | | |
| OH $\Pi_{3/2} - \Pi_{3/2}$ J = 7/2 – 5/2 | 84.5 | 120.61 | 59.8 | 5.5(0.3) | | | | | | | | | | |
| $H_2O$ $3_{22} - 2_{11}$ | 90.0 | 136.94 | 52.3 | 1.5(0.3) | | | | | | | | | | |
| $H_2O$ $2_{20} - 1_{11}$ | 101.0 | 53.44 | 39.3 | 4.4(0.2) | 313.7 | 2.5(0.2) | | | | | | | | |
| $H_2^{18}O$ $2_{20} - 1_{11}$ | 102.0 | 52.86 | 38.2 | 3.2(0.3) | | | | | | | | | | |
| $H_2O$ $2_{21} - 1_{10}$ | 108.1 | 60.96 | 32.3 | 3.2(0.3) | 269.2 | 2.7(0.3) | 345.0 | 1.8(0.2) | | | | | | |
| OH $\Pi_{3/2} - \Pi_{3/2}$ J = 5/2 – 3/2 | 119.3 | 0.04 | 23.1 | 14.0(0.4) | 222.7 | 5.3(0.3) | 265.5 | 4.2(0.2) | 58.7 | -1.7(0.2) | 306.6 | 2.1(0.2) | 41.9 | 3.6(0.5) |
| CH $\Pi_{3/2} - \Pi_{1/2}$ J = 3/2 – 1/2 | 149.2 | 0.07 | 10.7 | 3.5(0.3) | 109.4 | 4.9(0.3) | 137.0 | 2.1(0.1) | | | | | | |
| $OH^+$ $2_3 - 1_2$ | 153.2 | 46.60 | 9.7 | 7.2(0.3) | | | | | | | | | | |
| OH $\Pi_{1/2} - \Pi_{1/2}$ J = 3/2 – 1/2 | 163.3 | 181.82 | 7.6 | -6.3(0.4) | | | | | 98.4 | -2.3(0.1) | 22.6 | -2.3(0.2) | | |
| $H_2O$ $3_{03} - 2_{12}$ | 174.6 | 114.38 | 5.7 | 2.8(0.6) | | | | | | | | | | |
| $H_2O$ $2_{12} - 1_{01}$ | 179.5 | 34.19 | 4.8 | 10.3(0.8) | 64.0 | 4.9(0.6) | | | | | | | | |

NOTES. Column (1), lines are denoted by molecule and transition, Column (2), rest wavelength in microns, Column (3), lower level energy. Blended doublets are listed as a single transition with average values of $\lambda$ and $E_{lower}$. Columns (4 – 9), split columns

list continua in $10^{-15}$ W m$^{-2}$μm$^{-1}$, equivalent widths in $10^{-2}$ microns, and the equivalent width statistical uncertainties in $10^{-2}$ microns, in parentheses, for a Gaussian profile fit. A negative equivalent width indicates the line is in emission.

# Table A1

## Model Input Parameters And Description

| Type | Parameter | Assumed Value(s) |
|---|---|---|
| Ionizing Continuum | SED | AGN ($T=10^6$, $\alpha_{ox}=10^{-1.4}$, $\alpha_{UV}=10^{-0.5}$, $\alpha_x=10^{-1.0}$) Starburst (4 Myr continuous, power-law index of 2.35, star-formation rate of 1 $M_{sun}$ yr$^{-1}$, all other parameters set to Starburst99 default) |
| | Ionization parameter | $U = \dfrac{\phi_H}{n_H c}$, where $\phi_H$ is the flux of hydrogen ionizing photons, $n_H$ is the hydrogen density at the illuminated face (see above) and $c$ is the speed of light. Varies from $10^{-4}$ to $10^0$, in increments of 1 dex |
| | Cosmic-ray ionization rate ($\zeta_{cr}$) | $5 \times 10^{-17}$ s$^{-1}$ |
| Density | Hydrogen density at illuminated face of H$^+$ region, $n(H^+)$ | 30, 300, 3000 cm$^{-3}$ |
| | Equation of State[1] | Constant pressure (thermal, magnetic, and radiation). Assumption of constant pressure changes the hydrogen density with depth into the cloud, with no turbulent pressure other than a small turbulent broadening of 1 km s$^{-1}$ for its effects of line optical depths. |
| | Magnetic Field[2] | $B_{depth} = B_0 \left( \dfrac{n_{H(depth)}}{n_{H(initial)}} \right)^{\kappa}$, $B_0 = 100$ μG $n_H(initial) = n(H^+) = 30, 300,$ or $3000$ cm$^{-3}$, and $\kappa = 2/3$ |
| Abundances | Gas-phase abundances | ISM abundances as defined in *Cloudy*, which are taken from Cowie & Sonngaila (1986), Savage & Sembach (1996), Meyer et al. (1998). |
| | Dust Properties | "ISM grains", with $R_V = 3.1$ and $A_V/N(H_{tot}) = 5.4 \times 10^{-22}$ mag cm$^2$. Astronomical silicates and graphite grains are used (Martin & Rouleau 1991). There is a small variation (less than 5%) in $R_V$ and $A_V/N(H_{tot})$ with depth in each calculation due to variation in PAH abundance with depth (see below). |
| | PAH properties | Size distribution and abundance from Abel et al. (2008) and Paper I. PAH abundance scaled relative to atomic hydrogen fraction to mimic observed Similars across Orion Bar (Sellgren, Tokunaga, & Nakada 1990). |
| Structure | Geometry | 1D spherical, where gas fully covers the continuum source, as in Pellegrini et al. (2007) and Paper I |
| | Stopping condition | $N(H_{total}) = 10^{21.3}$ to $10^{24.9}$, in increments of 0.1 dex ($A_V = 1$ to 4000 mag) |

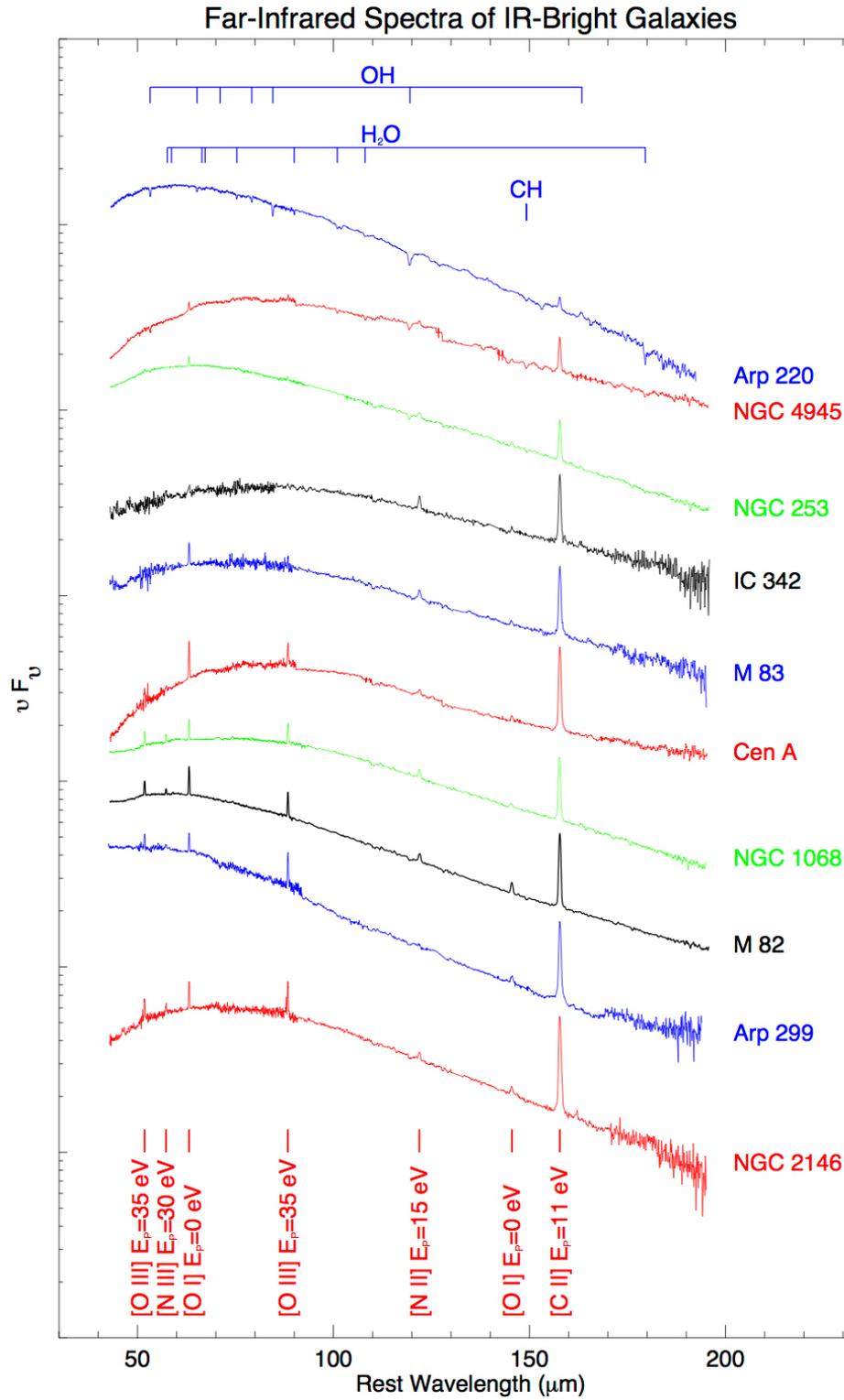

Figure 1. The full LWS spectra of the 10 IR-bright galaxies. The spectra have been shifted and ordered vertically, in order of increasing ([OIII] + [NIII]) / FIR, top to bottom.

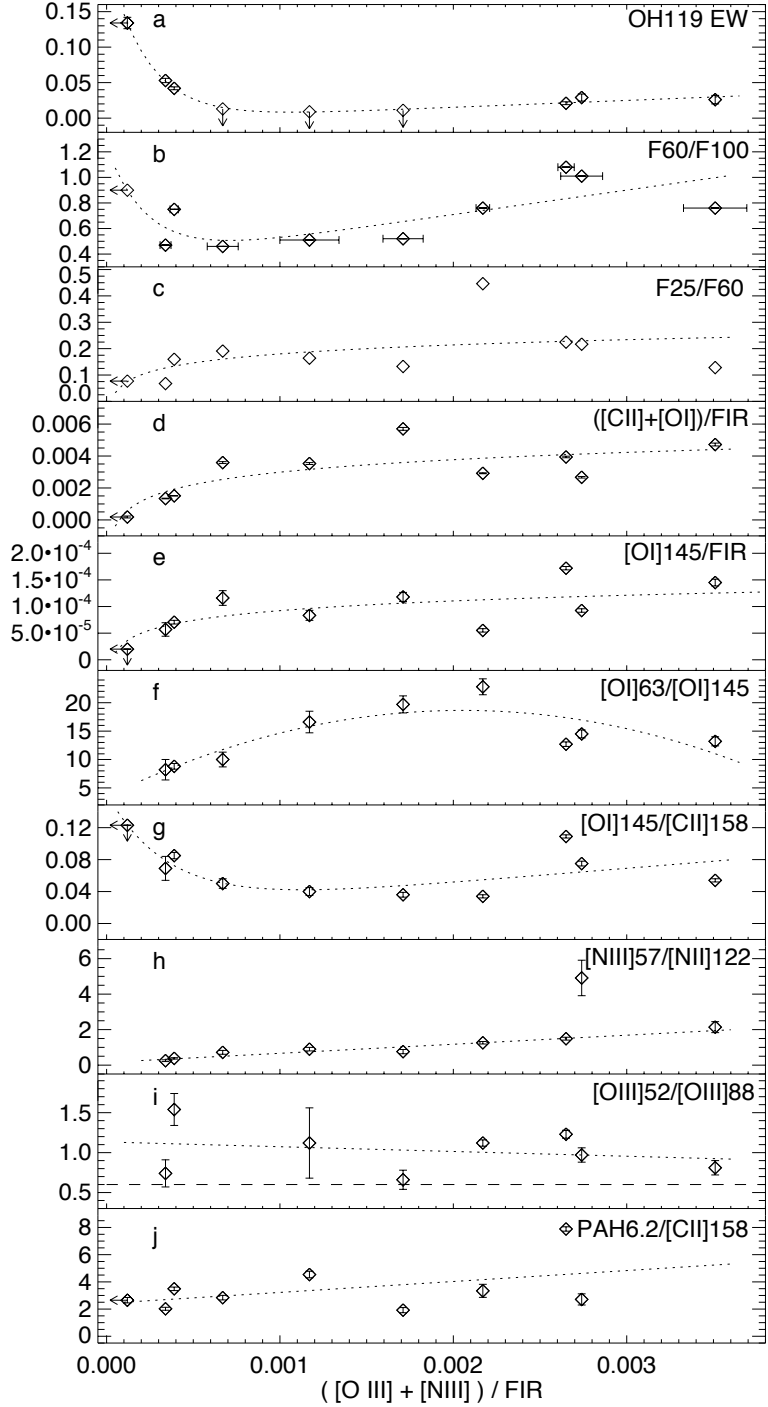

Figure 2. Plotted as a function of the ([OIII] + [NIII])/FIR flux ratio for the sample galaxies are the a) OH119 equivalent widths in microns, b),c) IRAS F60/F100 and F25/F60 flux ratios, d,e) the ([CII] + [OI])/FIR, [OI]145/FIR ratios, and f-j) [OI]63/[OI]145, [OI]145/[CII]158, [NIII]57/[NII]122, [OIII]52/[OIII]88, [PAH6.2/[CII]158 ratios, as labeled in the upper right corner of each panel. The dotted lines show fits to the data (see text) and the dashed line in panel i shows the theoretical [OIII] line ratio for the low density limit ($n \leq 100$ cm$^{-3}$). The error bars are based on propagation of the statistical 1-σ measurement uncertainties. ([OIII] + [NIII])/FIR error bars are displayed in panel b. Note that left-to-right corresponds to top-to-bottom in Figure 1.

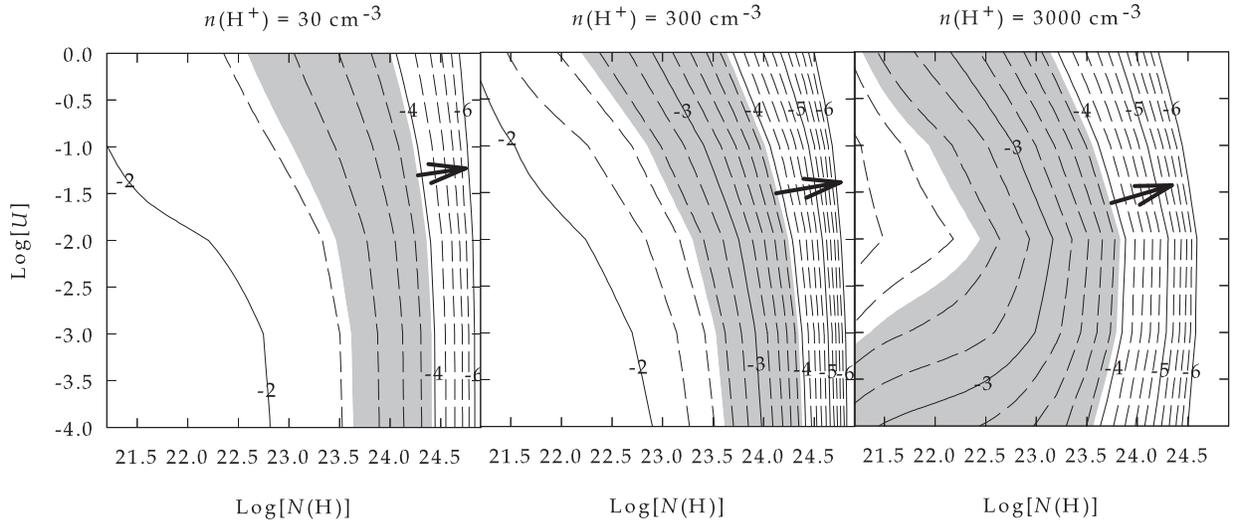

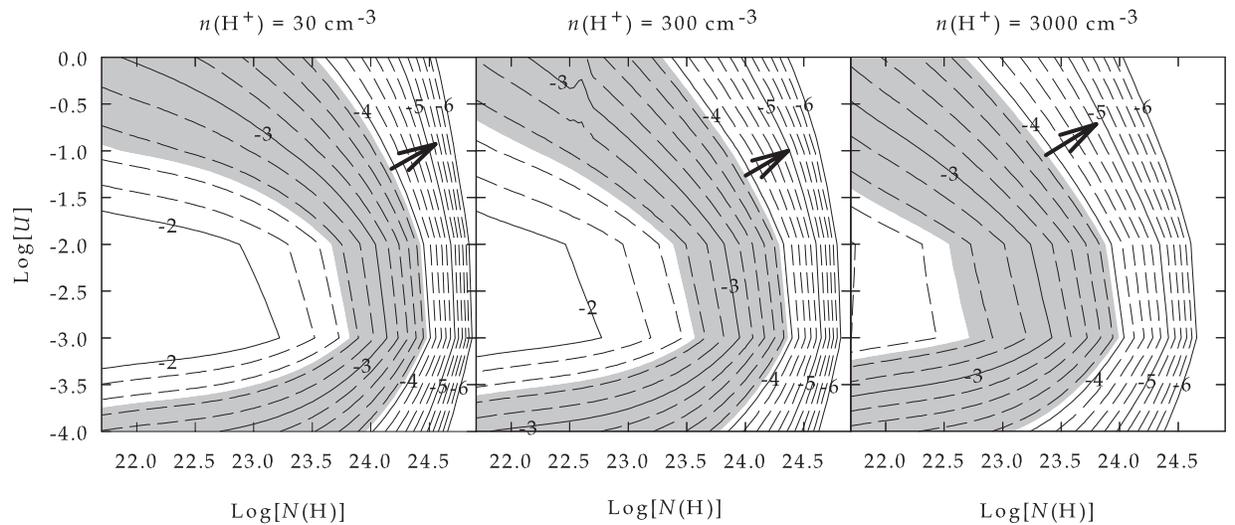

Figure 3a. Log [ ( [OIII] + [NIII]) / FIR ] contours for both starburst (upper) and AGN (lower) SEDs. The shaded regions mark the observed values for the sample galaxies, up to the upper limit measurement represented with arrows, in this case for Arp 220. (The length/angle of the arrows have no meaning.)

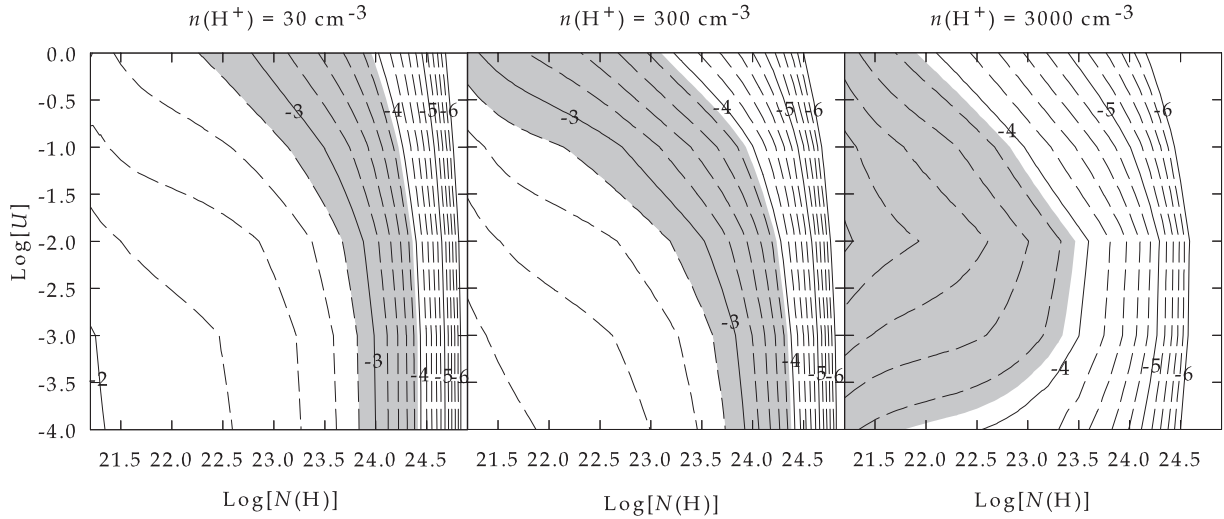

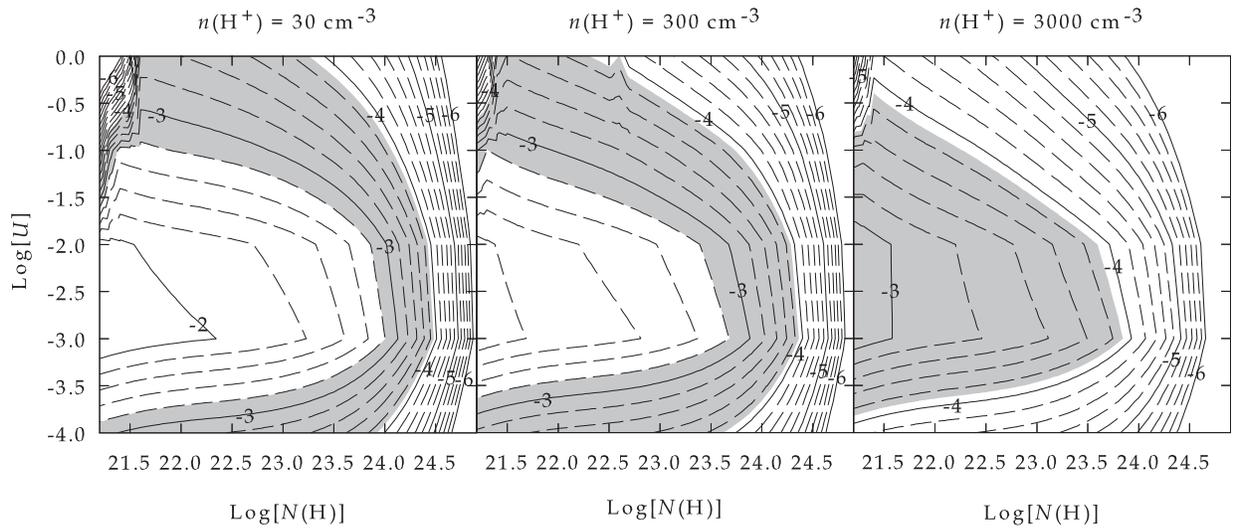

Figure 3b. Log [ [OIII]88 / FIR] contours for both starburst (upper) and AGN (lower) SEDs, shading as in Figure 3a.

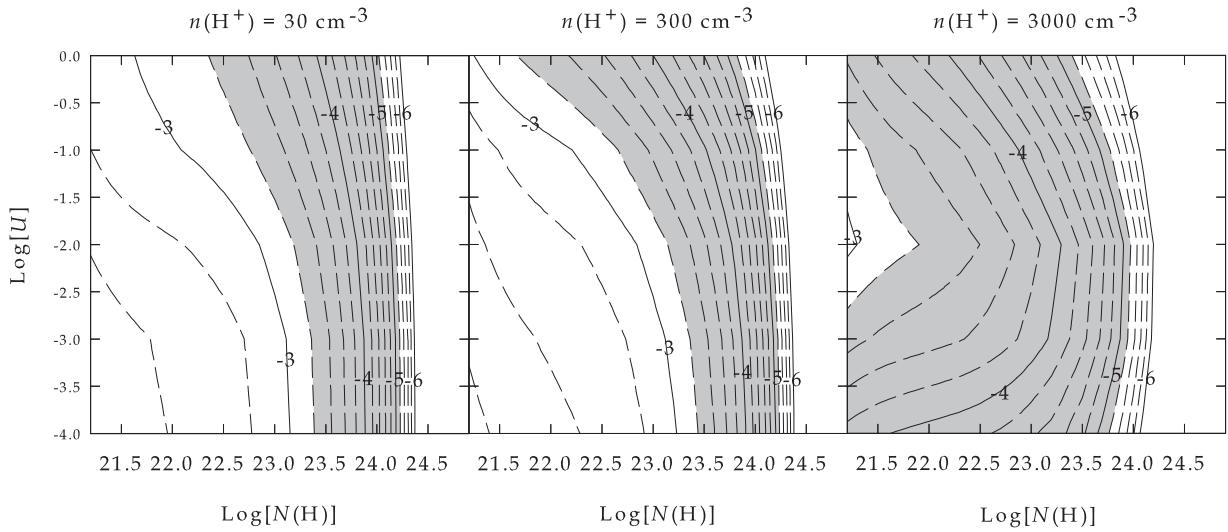

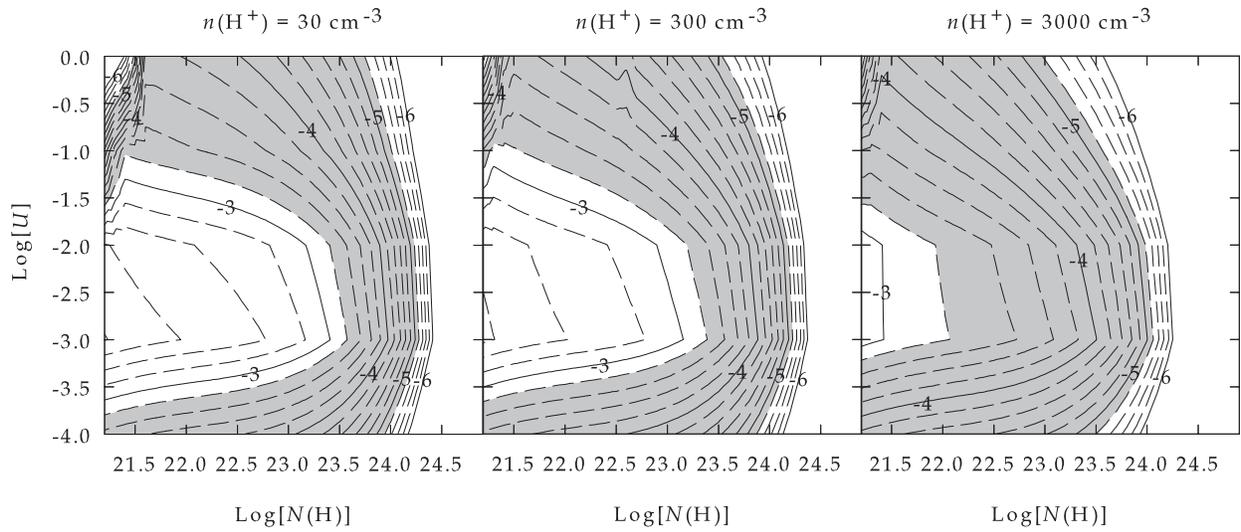

Figure 3c.  Log [ [NIII]57 / FIR] contours for both starburst (upper) and AGN (lower) SEDs, shading as in Figure 3a.

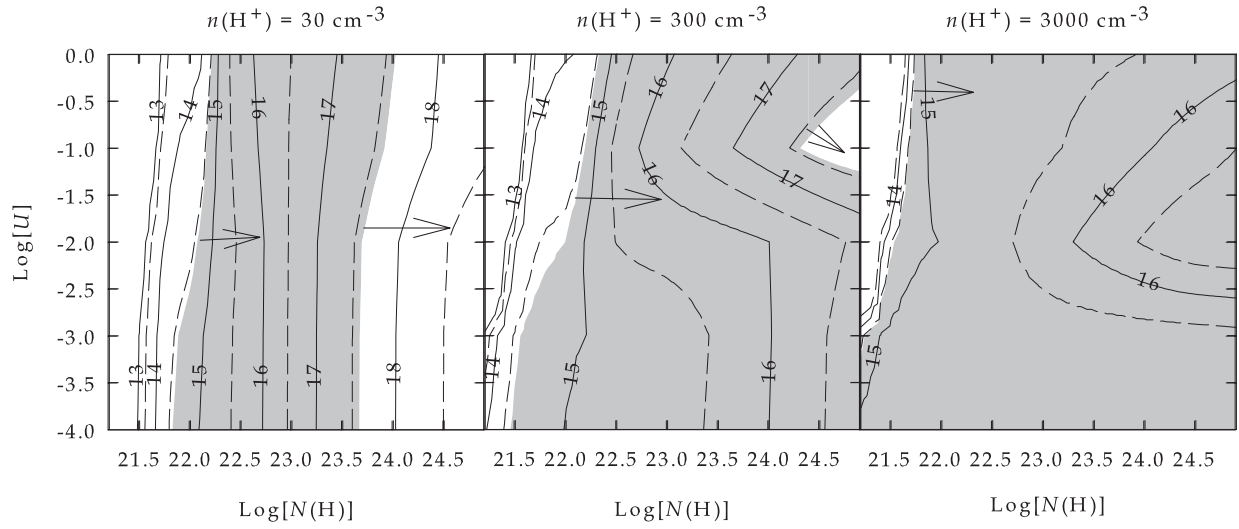

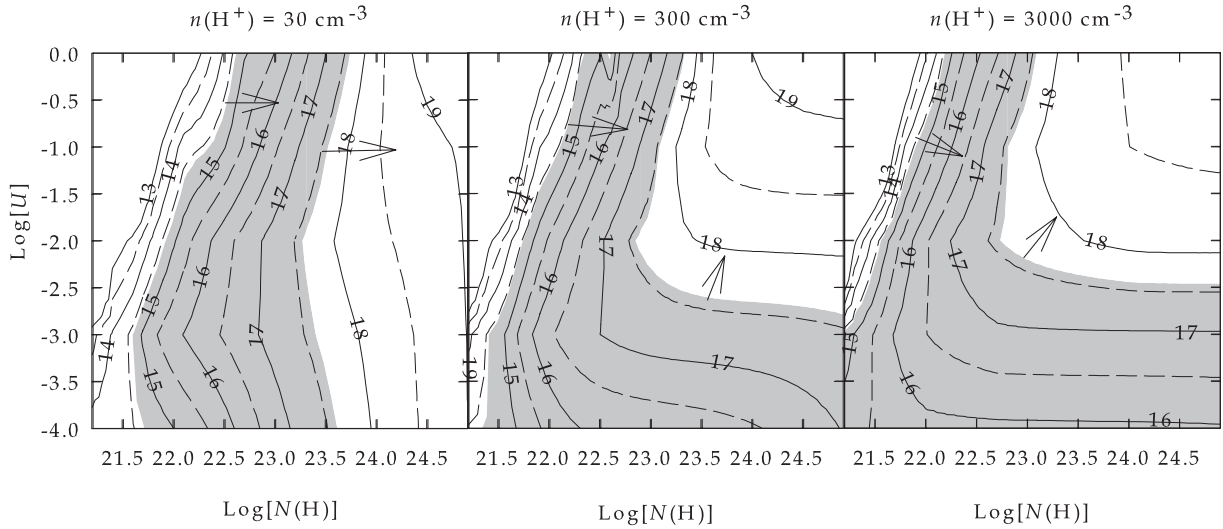

Figure 4a. Log [ $N$(OH)] contours for both starburst (upper) and AGN (lower) SEDs versus log $U$ and log $N$(H), shading and upper/lower limit arrows, as in Figure 3a.

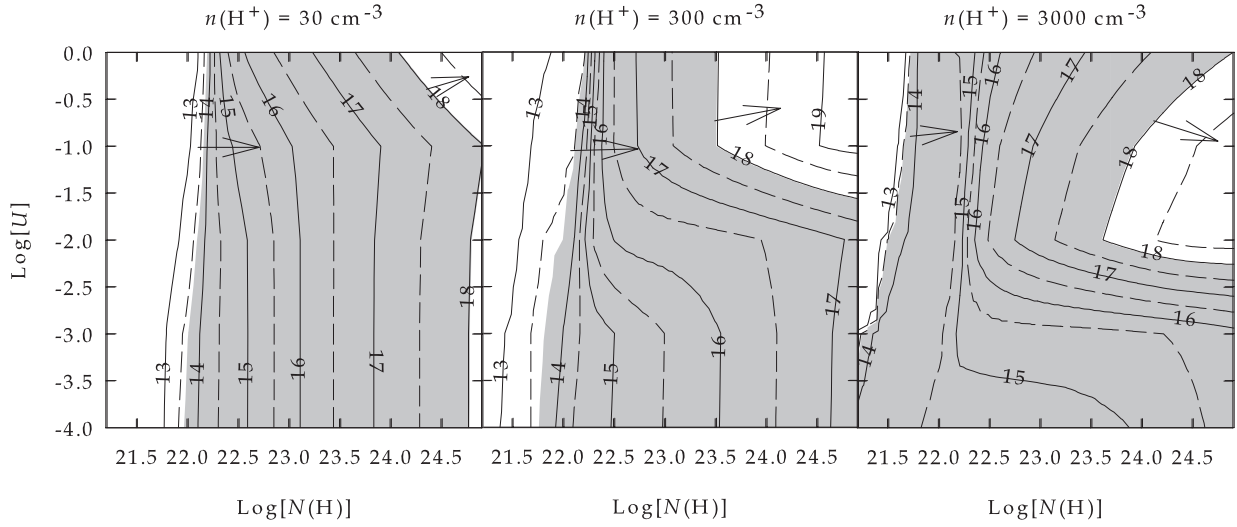

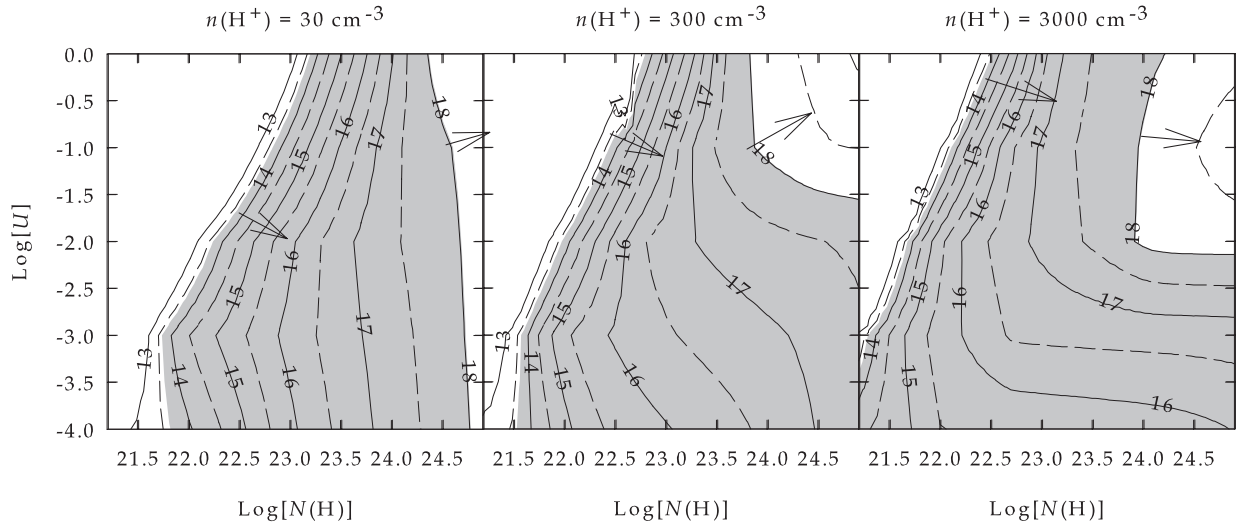

Figure 4b. Log [ $N(H_2O)$] contours for both starburst (upper) and AGN (lower) SEDs, shading and upper/lower limit arrows as in Figure 3a.

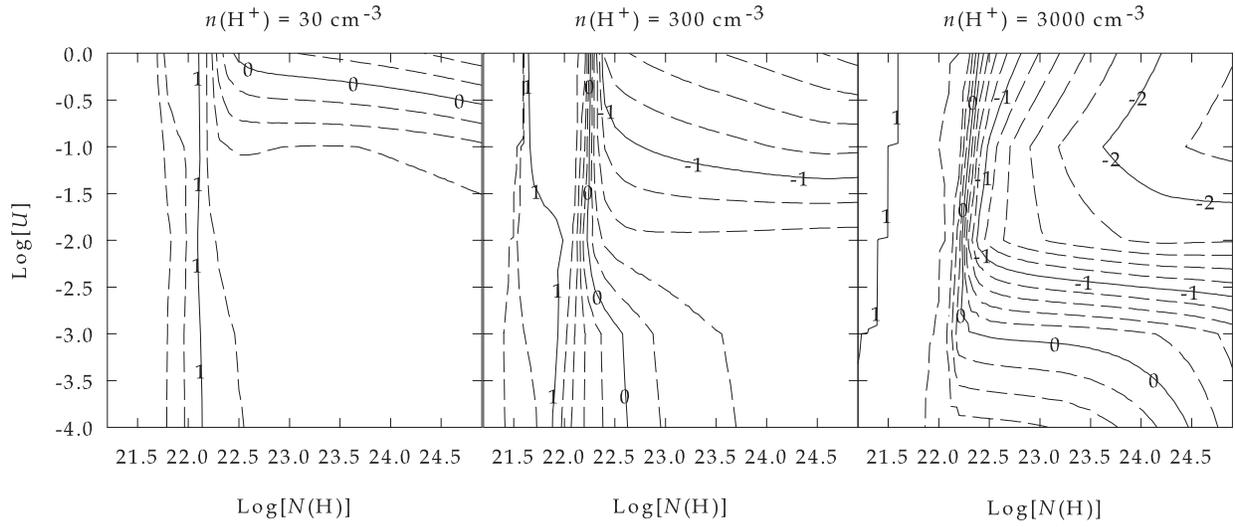
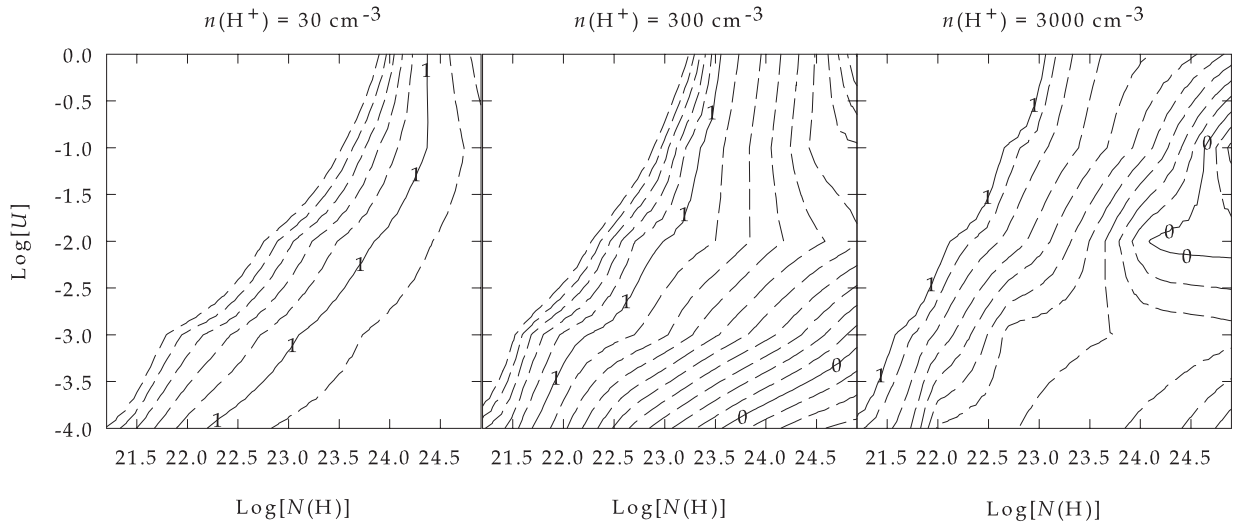

Figure 4c. Log [$N$(OH) / $N$(H$_2$O)] contours for both starburst (upper) and AGN (lower) SEDs.

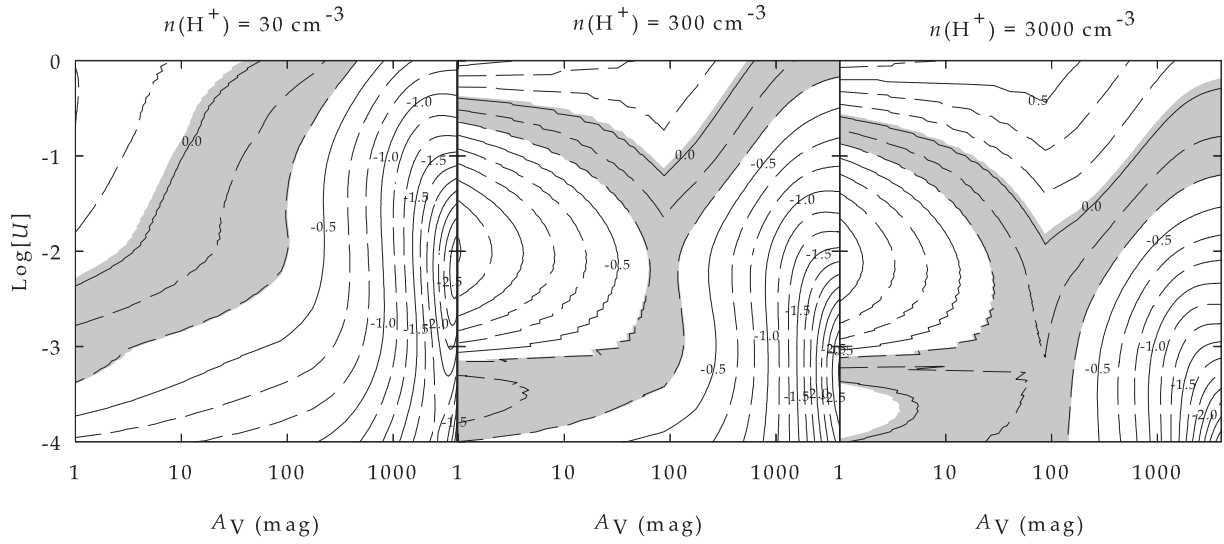

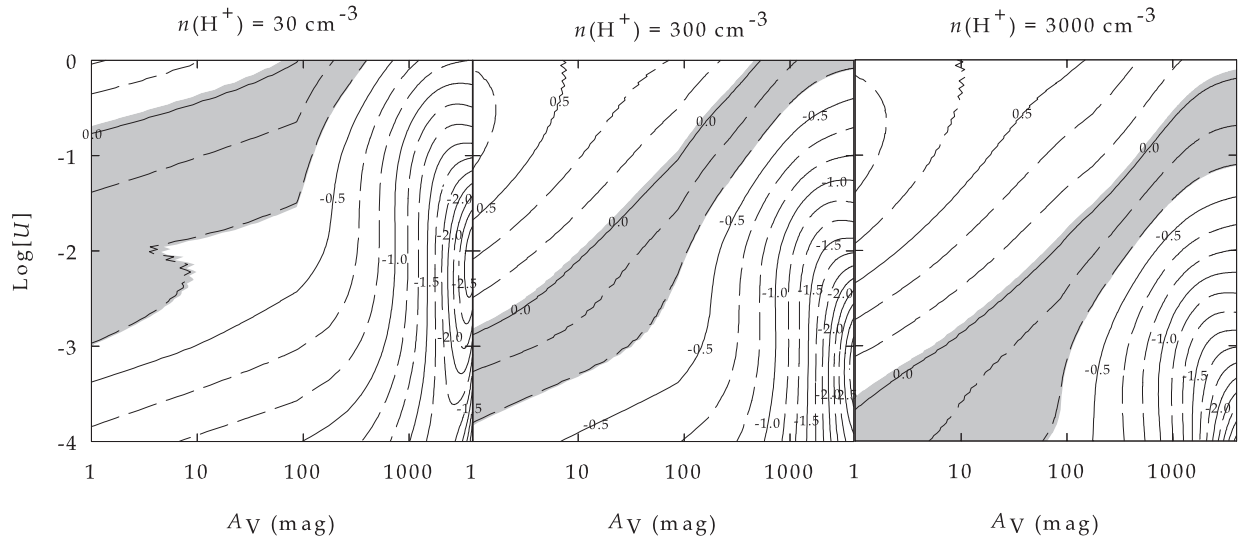

Figure 5. Log [ F60) / F100] contours for both starburst (upper) and AGN (lower) SEDs versus log $U$ and log $N$(H), shading as in Figure 3a.

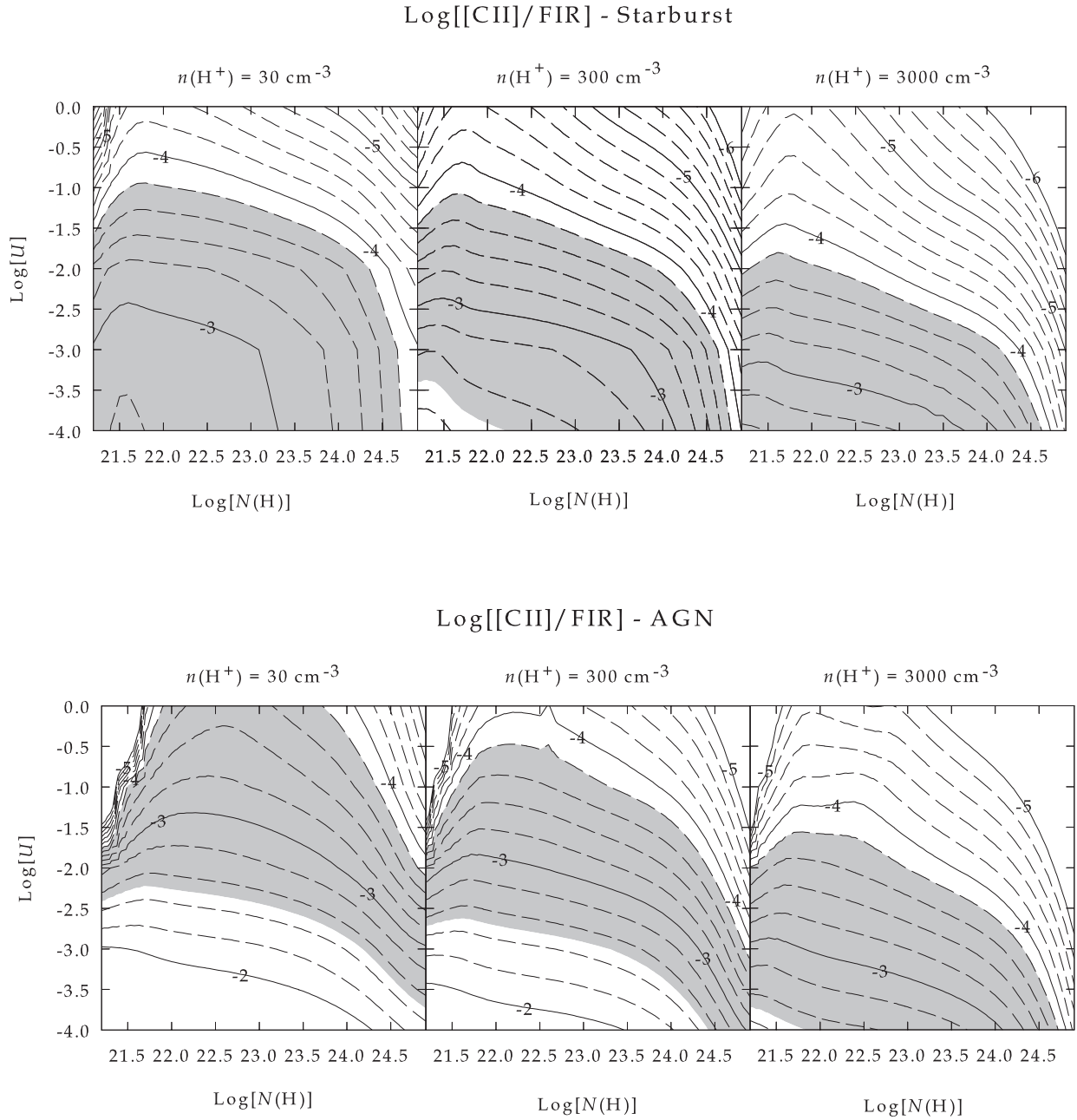

Figure 6a. Log [ [CII] / FIR] for both starburst (upper) and AGN (lower) SEDs as a function of log $U$ and log $N$(H), shading as in Figure 3a.

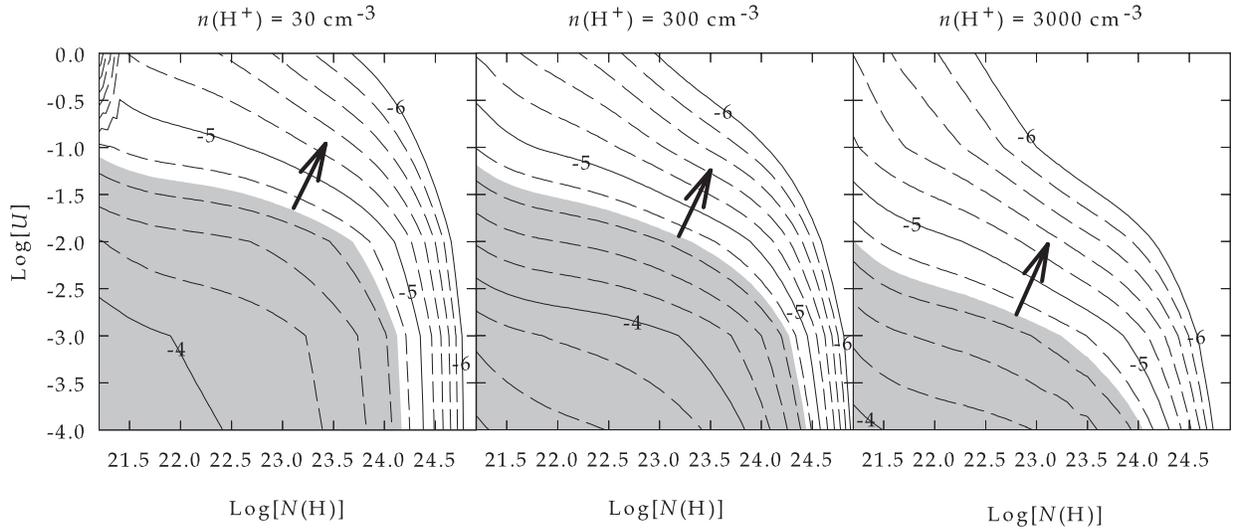

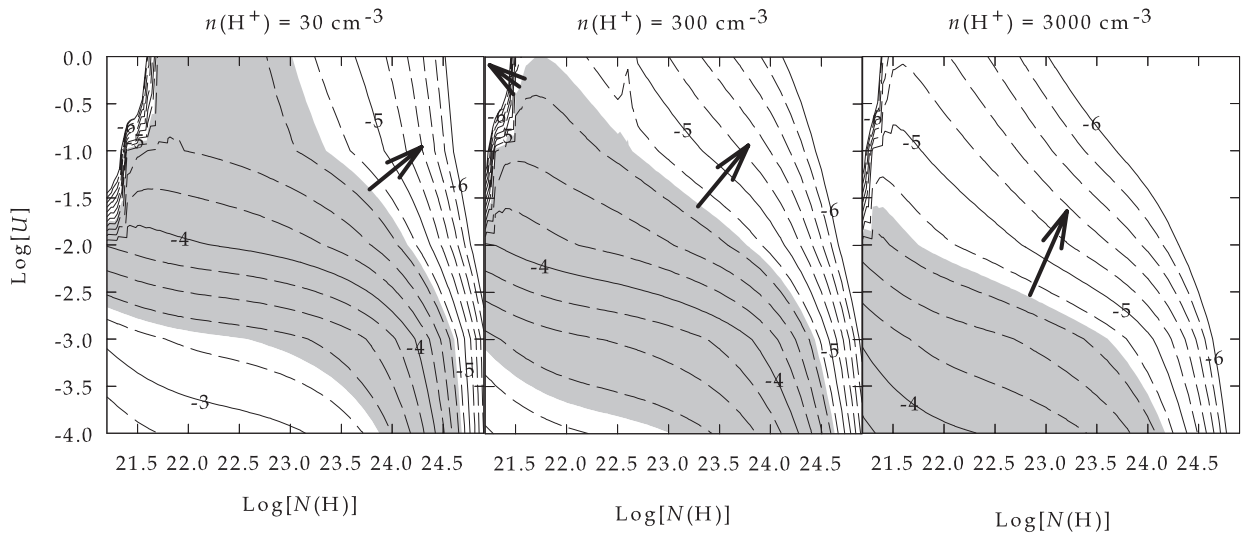

Figure 6b. Log [ [NII]122 / FIR] contours for both starburst (upper) and AGN (lower) SEDs, shading and upper limit arrows as in Figure 3a.

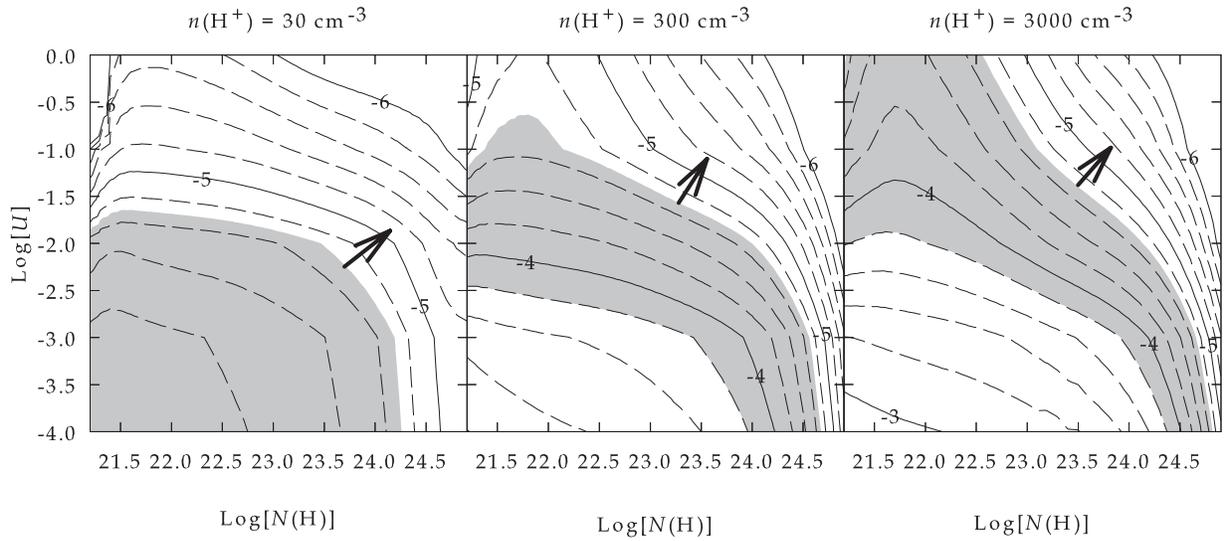

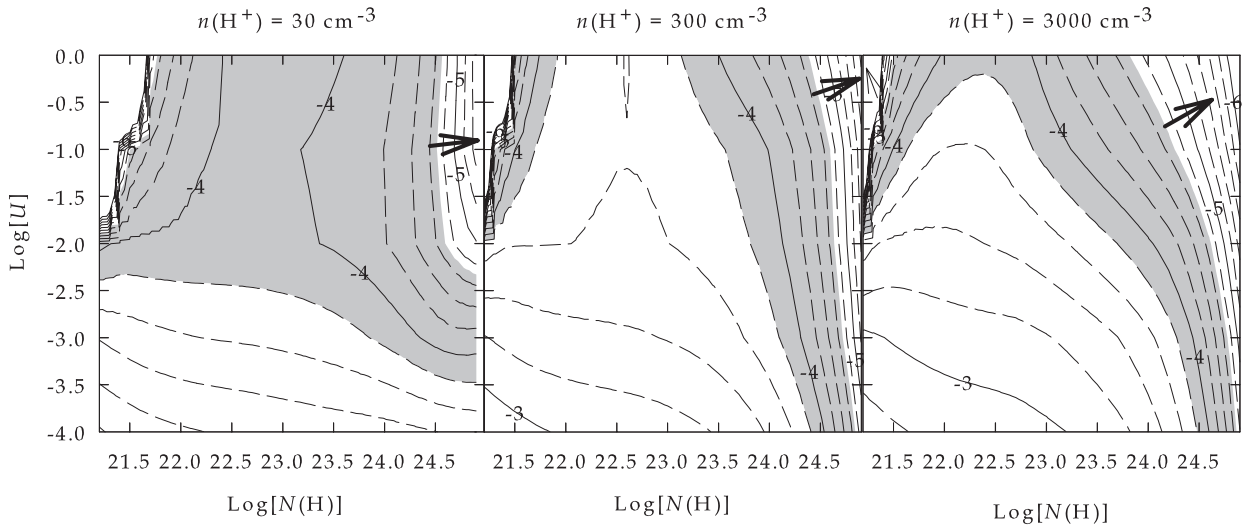

Figure 6c. Log [ [OI]145 / FIR] contours for both starburst (upper) and AGN (lower) SEDs versus log $U$ and log $N$(H), shading and upper limit arrows as in Figure 3a.

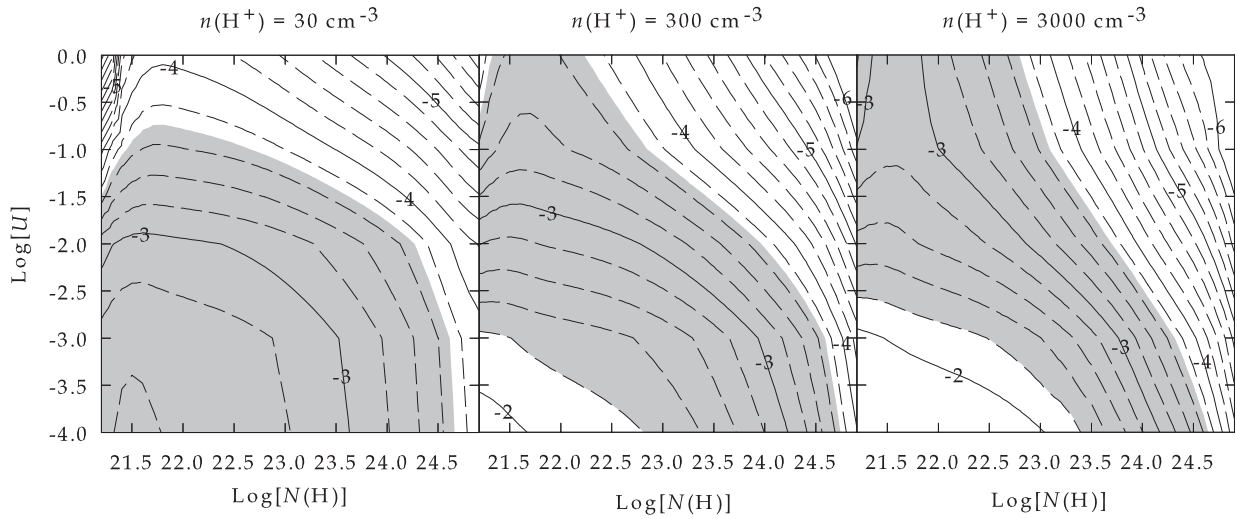

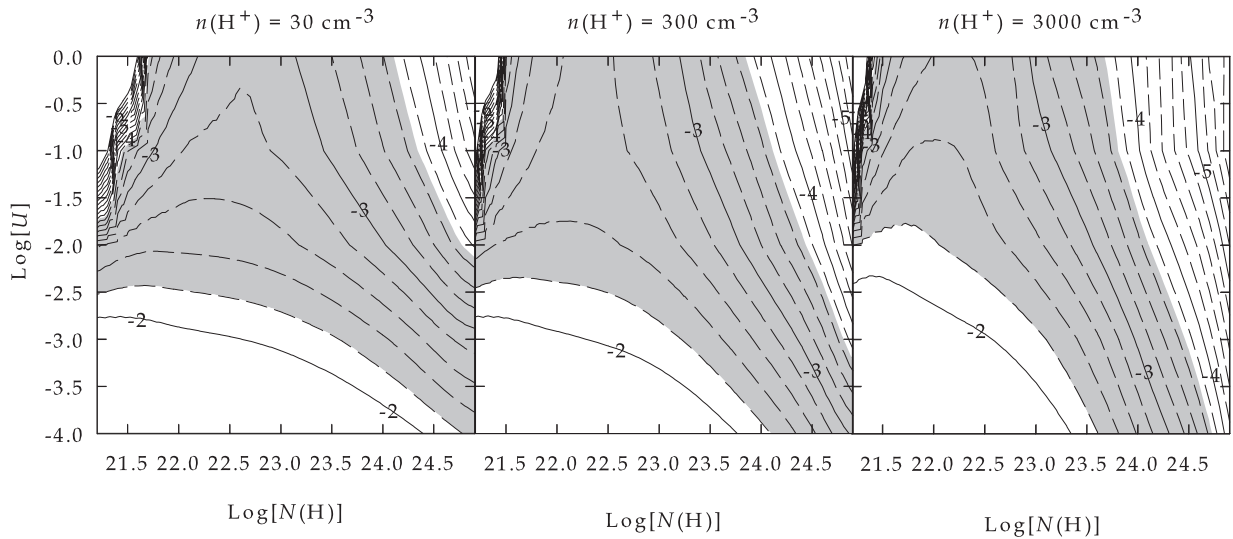

Figure 6d.  Log [ ( [CII]+[OI]145+[OI]63 ) / FIR] contours for both starburst (upper) and AGN (lower) SEDs, shading as in Figure 3a.

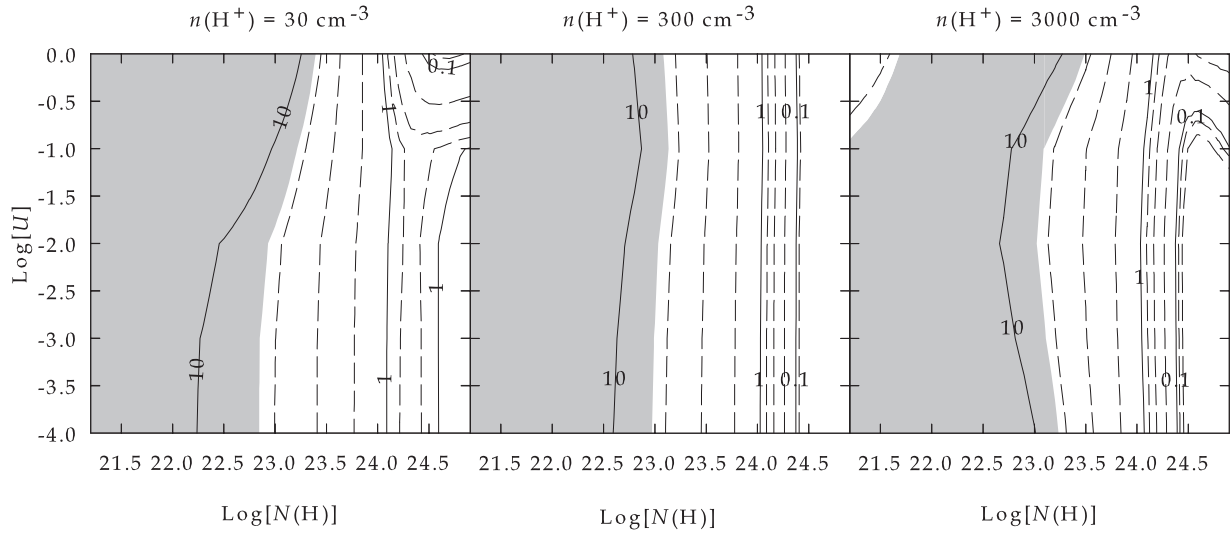

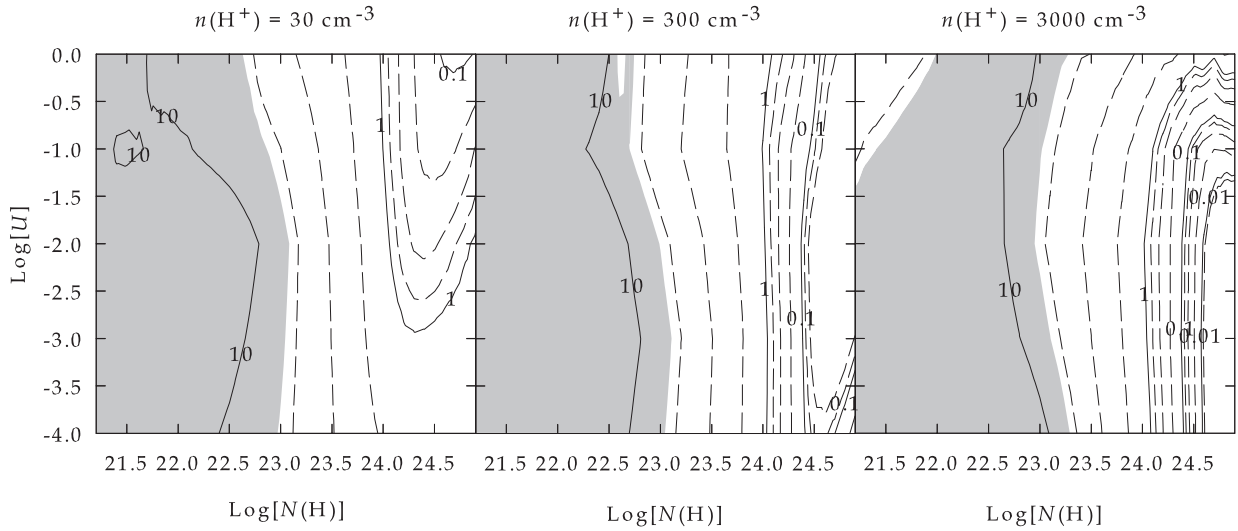

Figure 7a. [OI]63 / [OI]145 contours for both starburst (upper) and AGN (lower) SEDs, shading as in Figure 3a.

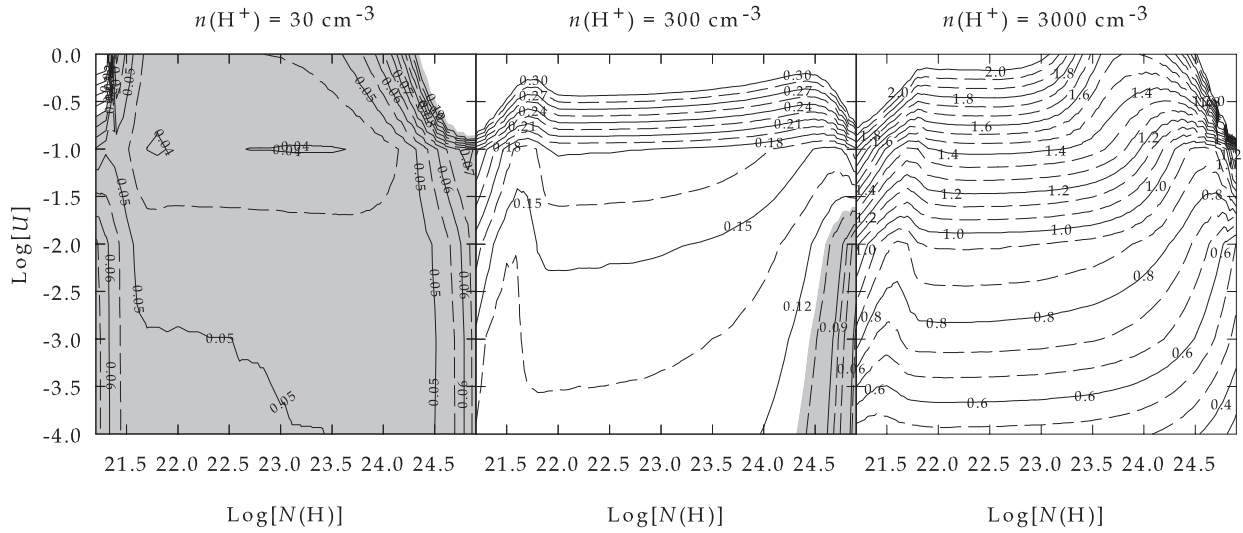

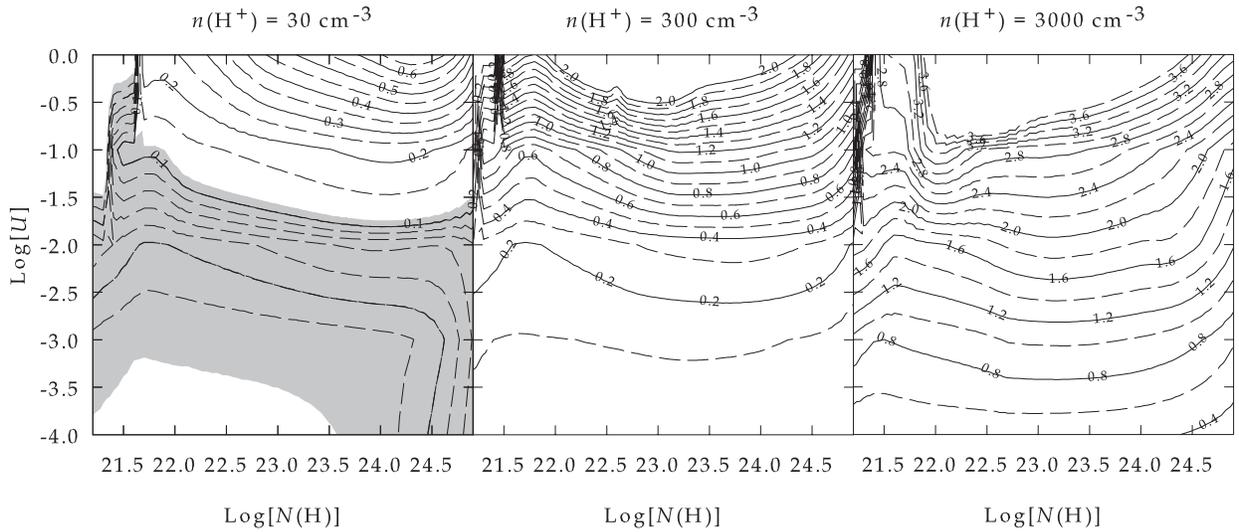

Figure 7b. [OI]146 / [CII]158 contours for both starburst (upper) and AGN (lower) SEDs versus log $U$ and log $N$(H), shading as in Figure 3a.

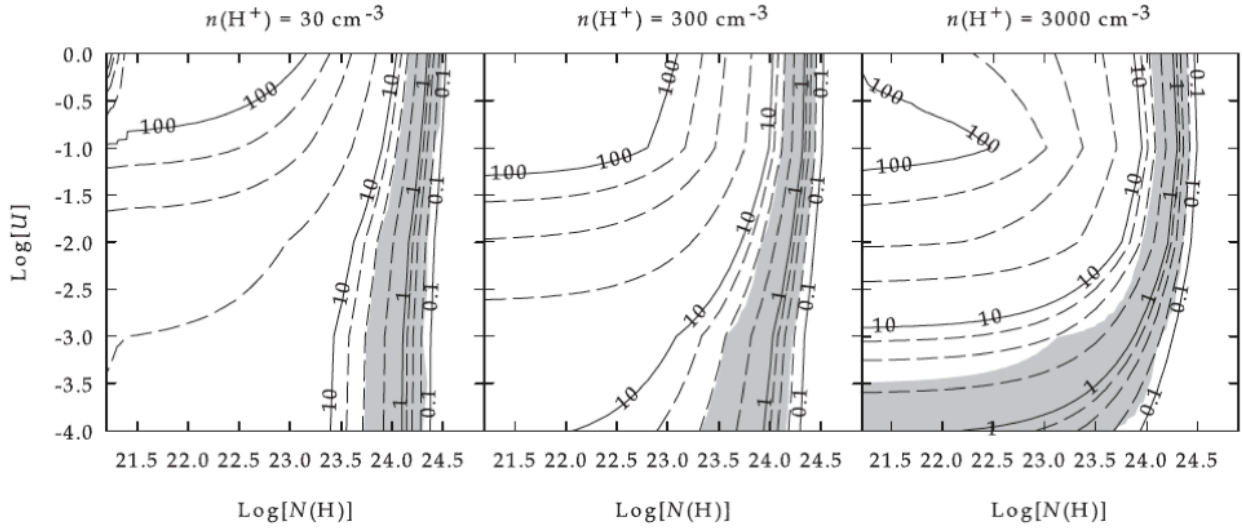

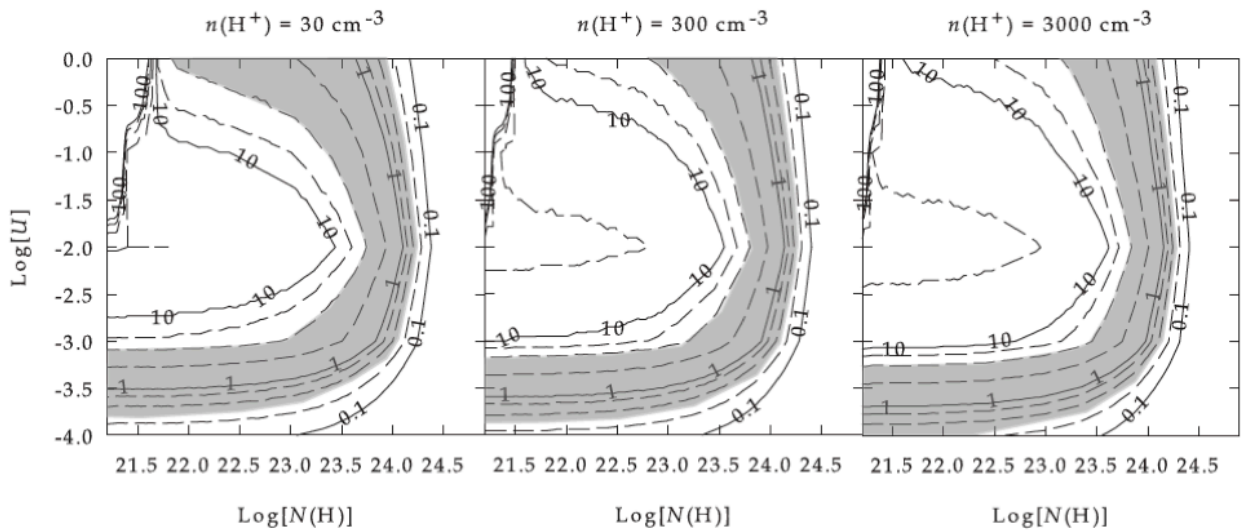

Figure 7c. [N III]57 / [N II]122 contours for both starburst (upper) and AGN (lower) SEDs, shading as in Figure 3a.

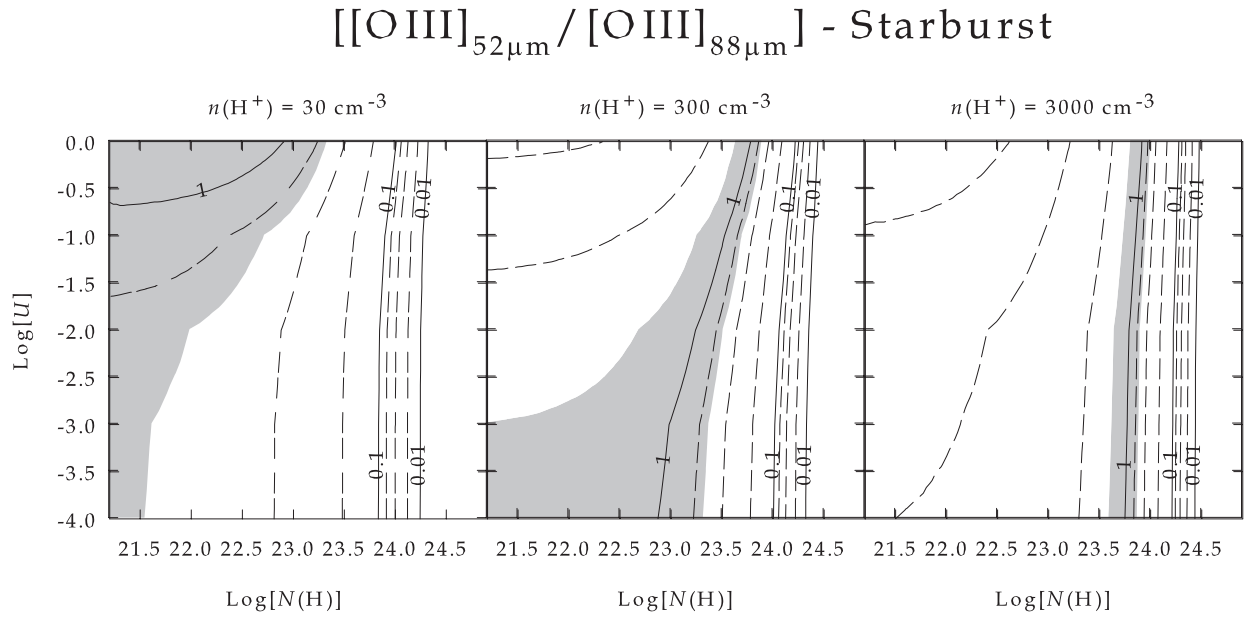

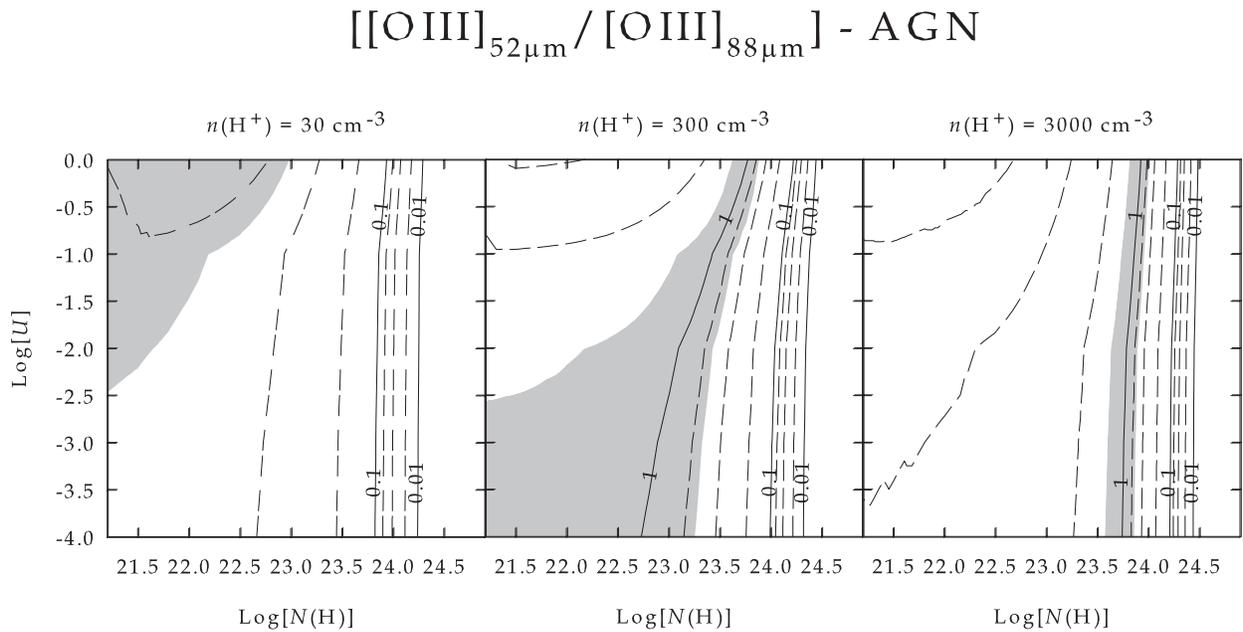

Figure 7d.  [OIII]52 / [OIII]88 contours for both starburst (upper) and AGN (lower) SEDs, shading as in Figure 3a.

51